\documentclass[preprint,12pt]{elsarticle}

\usepackage{amssymb}
\usepackage{amsmath}
\usepackage{amsthm}
\usepackage{bm}
\usepackage{braket}
\usepackage{caption}
\usepackage{hyperref}
\usepackage{float}
\usepackage[normalem]{ulem}
\usepackage{algorithm}
\usepackage{algorithmicx}
\usepackage{algpseudocode,pseudocode}
\hypersetup{
    colorlinks=true,
}

\newcounter{bla}

\journal{Computer Physics Communications}

\newcommand{\noi}{\noindent}
\newcommand{\mb}{\mathbf}

\newcommand\THEOSMARVEL{Theory and Simulation of Materials (THEOS) and National Centre for Computational Design and Discovery of Novel Materials (MARVEL), {\'E}cole Polytechnique F{\'e}d{\'e}rale de Lausanne, 1015 Lausanne, Switzerland}
\newcommand\UNIPD{Dipartimento di Fisica e Astronomia, Universit{\`a} di Padova, via Marzolo 8, I-35131 Padova, Italy}
\newcommand\Dublin{School of Physics and CRANN Institute, Trinity College Dublin, College Green, Dublin 2, Ireland}

\begin{document}

\begin{frontmatter}



\title{\texttt{SIMPLE} code: optical properties with optimal basis functions}


\author[a]{Gianluca Prandini\corref{author}}
\author[b,c]{Mario Galante}
\author[a]{Nicola Marzari}
\author[b]{Paolo Umari}

\cortext[author] {Corresponding author.\\\textit{E-mail address:} gianluca.prandini@epfl.ch}
\address[a]{\THEOSMARVEL}
\address[b]{\UNIPD}
\address[c]{\Dublin}

\begin{abstract}
We present \texttt{SIMPLE}, a code developed to calculate optical properties of metallic and insulating extended systems using the optimal basis method originally proposed by E. L. Shirley in 1996. 
Two different approaches for the evaluation of the complex dielectric function are implemented: (i) the independent-particle approximation considering both interband and intraband contributions for metals and (ii) the Bethe-Salpeter equation for insulators. Since, notably, the optimal basis set is systematically improvable, accurate results can be obtained at a strongly reduced computational cost. 
Moreover, the simplicity of the method allows for a straightforward use of the code; improvement of the optimal basis and thus the overall precision of the simulations is simply controlled by one (for metals) or two (for insulators) input thresholds.
The code is extensively tested, in terms of verification and performance, on bulk silver for metals and on bulk silicon for insulators.\\ 
\end{abstract}


\end{frontmatter}



\noindent
{\bf Program summary} \\

\begin{small}
\noindent
{\em Program Title:} \texttt{SIMPLE}   \\
{\em Licensing provisions:} GNU General Public License V 2.0 \\
{\em Programming language:} Fortran 95                         \\
{\em Distribution format:} tar.gz                         \\
{\em Computer:} Any computer architecture     \\
{\em External routines:} Quantum ESPRESSO distribution, BLAS, LAPACK, FFTW, MPI.     \\
{\em Nature of problem:} First-principles calculations of the optical properties of metals and insulators.     \\
{\em Solution method:} Shirley's optimal basis for the complex dielectric function, at the independent-particle level for metals and by solving the Bethe-Salpeter equation for insulators.    \\
{\em Restrictions:} Norm-conserving pseudopotentials.   \\

\end{small}

\section{Introduction}
\label{intro}

\noi
Density-functional theory (DFT) has largely proven itself to be a powerful and reliable tool to predict structural and vibrational properties (and, in general, any ground-state property) of complex materials. This is to be contrasted with well-known intrinsic limitations in the description of electronic excitations~\cite{Martin2004}. 
However, complementary first-principles methods for calculating optical properties  have been developed and applied to the study of materials, giving thus birth in recent years to the field of theoretical spectroscopies. In particular, the Bethe-Salpeter equation (BSE), based on many-body perturbation theory~\cite{Strinati1988, Onida2002, Martin2016}, has become the state-of-the-art approach for the simulation of optical spectroscopies, such as absorption or reflectivity, of extended systems.
Although usually at least one order of magnitude more demanding in terms of computational cost with respect to typical DFT simulations, the BSE approach  and the $GW$ method~\cite{Hybertsen1986}  have proven to be the most accurate and predictive methodologies for the study of neutral and charged excitations in extended systems, respectively (we mention in passing the development of Koopmans' compliant spectral functionals as an alternative to $GW$~\cite{Dabo2010, Borghi2014, Nguyen2018, Colonna2018}). 
Leveraged by the computational power available nowadays and due to the large range of technologically relevant applications, such as photovoltaics, plasmonics and optoelectronics, first-principles methods for the calculation of optical properties are starting to be routinely used not just to help analyzing and interpreting experimental data but also to guide the discovery of novel materials by computational design~\cite{Marzari2016}. 
This trend is evident from the fact that methods for theoretical spectroscopies are now being implemented and becoming available to researchers in different specialized codes, such as Yambo~\cite{Marini2009} or BerkeleyGW~\cite{Deslippe2012} (see Ref.~\cite{Leng2016} for a more comprehensive list), and interfaced with broadly available DFT engines (e.g. Quantum ESPRESSO~\cite{Giannozzi2009}, Abinit~\cite{Gonze2009} and Siesta~\cite{Soler2002} or directly incorporated in these~\cite{Kresse1996, Hutter2014, Gonze2016, Giannozzi2017}).\\
We present here the \texttt{SIMPLE} code, which has been implemented as a component of the Quantum ESPRESSO (QE) distribution.
The main purpose of the code is to provide accurate and, at the same time, efficient first-principles simulations of the optical properties of both insulating and metallic materials. For metals, where the electron-hole interaction is screened by conduction electrons, the independent-particle (IP) approximation including both interband and intraband contributions has been implemented. For insulators instead, where the inclusion of the electron-hole interaction is often compulsory to correctly simulate optical spectra, the BSE approach has been implemented.
In order to increase the efficiency of the calculation of the IP and BSE optical properties we exploit the method originally developed by E. L. Shirley in 1996~\cite{Shirley1996}. It is based on the use of an optimal basis (OB) set to represent the periodic part of the Bloch wavefunctions, and in this work we introduce a generalization of the method to represent also products of periodic functions. 
The most attractive feature of this approach resides in the fact that it allows a significant reduction of the computational cost of the simulations by introducing a well-controlled numerical error.\\
The paper is organized as follows: in Section 2 we introduce the most important theoretical concepts deployed in the code. In Section 3 we describe the main algorithms implemented, in particular the OB method and its application and extension to calculate optical properties of metals at the IP level and of insulators by solving the BSE. Then in Section 4 the structure of the \texttt{SIMPLE} code is presented and the main input parameters are described in detail. Finally in Section 5 and Section 6 the verification and the performance of the \texttt{SIMPLE} code are discussed.\\


\section{Theoretical background}
\label{theo_back}

\noindent
A natural physical quantity to study in first-principles simulations of optical properties is the full dielectric function, a complex (as opposite to real) dynamical (i.e. frequency-dependent) quantity. In the following we will show the basic equations that relate this macroscopic property of a material to its underling microscopic electronic structure, both within the framework of the IP approximation and of the BSE approach. Throughout the paper we use Hartree atomic units ($\hbar=m=e^2=1$) and Gaussian units ($4\pi\varepsilon_0=1$) for electromagnetism.\\
The general expression of the IP  dielectric function\footnote{
Here we neglect the fact that the dielectric function is actually a $3\times 3$ complex tensor whose additional off-diagonal terms describe subtle magnetic phenomena beyond the scope of this work, such as the Kerr effect or the anomalous Hall effect~\cite{Yao2004, Sangalli2012}.
In the following we will deal exclusively with the diagonal terms which are specified by the direction of the transferred momentum $\mb{q}$.} (valid both for metals and insulators) at a generic transferred momentum $\mb{q}$ and frequency $\omega$ is~\cite{Sangalli2017}
\begin{equation} \label{eq: eps_IP}
\varepsilon_{\text{IP}}(\mb{q},\omega) = 1  - \frac{4\pi}{|\mb{q}|^2} \frac{1}{V}\sum_{\mb{k}} \sum_{n , n'} \left( f_{n\mb{k}} - f_{n'\mb{k}+\mb{q}} \right) 
\frac{| \bra{\psi_{n'\mb{k}+\mb{q}}}  e^{i\mb{q}\cdot \mb{r}} \ket{\psi_{n\mb{k}}}  |^2 }{ \omega - ( E_{n'\mb{k}+\mb{q}} -  E_{n\mb{k} } )  
 +  i\eta  }
\end{equation}


\noindent
which is written in terms of the solutions of the one-particle Schr\"{o}dinger equation, $H^{\text{KS}}\ket{\psi_{n\mb{k}}}=E_{n\mb{k} }\ket{\psi_{n\mb{k}}}$, where $H^{\text{KS}}$ is the Kohn-Sham (KS) Hamiltonian. $f_{n\mb{k}}$ gives the occupation according to the Fermi-Dirac distribution of the KS Bloch state $\ket{\psi_{n\mb{k}}}$ with band index $n$ and wavevector $\mb{k}$ while $V$ is the volume of the crystal.
Instead, $\eta$ is an infinitesimal broadening introduced to perform the adiabatic switching-on of the perturbation and which, in practice, is used as an empirical broadening with the purpose to account for scattering processes, always present in real materials, and/or for finite experimental resolution.
\\
We recall here that the Bloch wavefunctions have the form  
\begin{equation}
\psi_{n\mb{k}}(\mb{x})=e^{i\mb{k}\cdot \mb{r}}u_{n\mb{k}}(\mb{x}),
\end{equation}

\noi
where $\mb{x}$ is a combined position and spin coordinate, $\mb{x}=(\mb{r},\sigma)$, and the function $u_{n\mb{k}}(\mb{x})$ is periodic with the periodicity of the crystal. In the non-collinear case, where spin-orbit interaction is explicitly included in the DFT calculation, $n$ has to be understood as a spinorial band index and $\ket{\psi_{n\mb{k}}}$ as a two-component spinor.\\
On the other hand the dielectric function obtained from the solution of the BSE within the Tamm-Dancoff approximation is given by an expression~\cite{Gatti2013} (valid for finite-gap systems) very similar to the IP case of Eq. \ref{eq: eps_IP}
\begin{equation} \label{eq: eps_BSE}
\varepsilon_{\text{BSE}}(\mb{q},\omega) = 1 -
\frac{4\pi}{|\mb{q}|^2} \frac{1}{V} 
\sum_{\lambda} \frac{ \left| \sum_{v,c,\mb{k}} \bra{\psi_{v\mb{k}+\mb{q}}}  e^{i\mb{q}\cdot \mb{r}} \ket{\psi_{c\mb{k}}} A_{\lambda,\mb{q}}^{(vc\mb{k})} \right|^2 }{ \omega - \tilde{E}_{\lambda}(\mb{q}) + i\eta },
\end{equation}

\noi
where the broadening $\eta$ has to be interpreted as in Eq.~\ref{eq: eps_IP}.
The main difference with respect to the IP expression is that Eq.~\ref{eq: eps_BSE} is written in terms of the solutions of an effective two-particle Schr\"{o}dinger-like equation 
\begin{equation}
\label{EQ:excitonic_hamiltonian}
H^{\text{exc}}_{vc\mb{k},v'c'\mb{k}'}(\mb{q})A^{(v'c'\mb{k}')}_{\lambda,\mb{q}}=\tilde{E}_{\lambda}(\mb{q}) A^{(vc\mb{k})}_{\lambda,\mb{q}},
\end{equation}

\noi
where $H^{\text{exc}}_{vc\mb{k},v'c'\mb{k}'}(\mb{q})$ is the excitonic Hamiltonian matrix\footnote{In Eq. \ref{eq: eps_BSE} it is understood that the long-range component of the bare Coulomb interaction in the excitonic Hamiltonian is not included.} in the basis of the single-particle Bloch wavefunctions of electron and hole states.
In this basis, $A_{\lambda,\mb{q}}^{(vc\mb{k})}= \braket{\psi_{v\mb{k}+\mb{q}} \psi_{c\mb{k}} |A_{\lambda,\mb{q}} }$ are the excitonic amplitudes labelled by the quantum numbers $(\lambda,\mb{q})$ and having energy $\tilde{E}_{\lambda}(\mb{q})$\footnote{
We have introduced the notation used throughout the paper $$\braket{ \mathbf{x}|\psi_{v\mb{k}+\mb{q}} \psi_{c\mb{k}}} = \psi_{v\mb{k}+\mb{q}} \left(\mathbf{x}\right) \psi_{c\mb{k}}\left(\mathbf{x}\right)$$, where the indeces $v$ and $c$ stand for valence band and conduction band, respectively.}. 
We note in passing that both for Eq.~\ref{eq: eps_IP} and Eq.~\ref{eq: eps_BSE} $GW$ energies (at the $G_0W_0$ level) or even eigenvectors (e.g. at the self-consistent $GW$ level) could be substituted for the KS ones; this is often the practice, especially for Eq.~\ref{eq: eps_BSE}, albeit at a largely increased computational cost. \\
We are interested in studying the optical limit, $\mb{q}\to 0$, for which the transferred momentum of the photons is negligible with respect to the size of the Brillouin zone (BZ). In general the dielectric function depends on the direction $\hat{\mb{q}} = \mb{q}/|\mb{q}|$ of the perturbing electric field and only for crystals with cubic symmetry the dielectric tensor is equivalent to a scalar. 
In the following we will study the optical limit of Eq.~\ref{eq: eps_IP} and Eq.~\ref{eq: eps_BSE} and we will keep explicit the dependence on the direction $\hat{\mb{q}}$ for the general case of anisotropic systems.



\subsection{IP}
\noindent
In metals, the IP dielectric function of Eq.~\ref{eq: eps_IP} in the optical limit can be divided into two separate contributions, an intraband Drude-like term due to the conduction electrons at the Fermi surface and an interband term due to vertical transitions between occupied and unoccupied bands~\cite{Wooten1972,Marini2001, Harl_phd}
\begin{equation} \label{eq: eps_IP_inter+intra}
\varepsilon_{\text{IP}}(\hat{\mb{q}},\omega) = \varepsilon_{\text{IP}}^{\text{inter}}(\hat{\mb{q}},\omega) + \varepsilon_{\text{IP}}^{\text{intra}}(\hat{\mb{q}},\omega) ,
\end{equation}
\noi
where
\begin{align} \label{eq: eps_IP_inter}
\varepsilon_{\text{IP}}^{\text{inter}}(\hat{\mb{q}},\omega) &= 1  - \frac{4\pi}{V} \sum_{\mb{k}} \sum_{n\neq n'} 
\frac{| \bra{\psi_{n'\mb{k}}}  \hat{\mb{q}} \cdot \mb{v} \ket{\psi_{n\mb{k}}}  |^2 }{(E_{n'\mb{k}} -  E_{n\mb{k}})^2}
\frac{ f_{n\mb{k}} - f_{n'\mb{k}}   }{ \omega - ( E_{n'\mb{k}} -  E_{n\mb{k} } )  
 +  i\eta   },  \\ \label{eq: eps_IP_intra}
\varepsilon_{\text{IP}}^{\text{intra}}(\hat{\mb{q}},\omega) &=  - \frac{\omega^2_{\text{D}}(\hat{\mb{q}})}{\omega(\omega + i\gamma)}.
\end{align}

\noi
We have defined the Drude plasma frequency as
\begin{equation} \label{eq: drude_plasma_freq}
\omega^2_{\text{D}}(\hat{\mb{q}})  = \frac{4\pi}{V} \sum_{\mb{k}} \sum_{n} | \bra{\psi_{n\mb{k}}}  \hat{\mb{q}} \cdot \mb{v} \ket{\psi_{n\mb{k}}}  |^2  \left(  -\frac{\partial f}{\partial E } \right),
\end{equation}

\noi
where $\mb{v}=  -i \, [\mb{r},H^{\text{KS}}]$ is the velocity operator and $\gamma$ is an empirical broadening representing dissipation effects of the conduction electrons, whose value can be taken from experimental data or estimated from additional first-principles calculations~\cite{Mustafa2016}. In the limit of zero temperature, the derivative of the Fermi-Dirac distribution in Eq.~\ref{eq: drude_plasma_freq} is a Dirac-delta function centred at the chemical potential and the Drude plasma frequency $\omega_{\text{D}}(\hat{\mb{q}})$ is always zero in finite-gap systems.

\subsection{BSE} 
\noi
The expression in Eq.~\ref{eq: eps_BSE} of the BSE dielectric function, in the optical limit $\mathbf{q}\rightarrow 0$, is given by~\cite{Albrecht}
\begin{equation} \label{eq: eps_BSE_optic}
 \varepsilon_{\text{BSE}}(\hat{\mb{q}},\omega)= 1 -
 \frac{4\pi}{V} \sum_{\lambda} \frac{1}{\omega - \tilde{E}_{\lambda} + i\eta} 
 \left| \sum_{v,c,\mb{k}} \frac{ \bra{\psi_{v\mb{k}}} \hat{\mb{q}} \cdot \mb{v} \ket{\psi_{c\mb{k}}} }{ E_{c\mb{k}} - E_{v\mb{k}}  } A_{\lambda}^{(vc\mb{k})} \right|^2,
\end{equation}

\noi 
where we construct the excitonic Hamiltonian considering only vertical electron-hole transitions ($\mb{q}=\mb{0}$). This means that $\tilde{E}_{\lambda}$ and $A_{\lambda}^{(vc\mb{k})}$ in Eq.~\ref{eq: eps_BSE_optic} are the eigenvalues and eigenfunctions of the Hamiltonian $H^{\text{exc}}_{vc\mb{k},v'c'\mb{k}'} = H^{\text{exc}}_{vc\mb{k},v'c'\mb{k}'}(\mb{q}=\mb{0})$.   

\noi
The excitonic Hamiltonian at $\mb{q}=0$ can be divided into three different contributions~\cite{Rohlfing2000}:
\begin{equation} \label{eq: H_exc_vck}
H^{\text{exc}}_{vc\mb{k},v'c'\mb{k}'} = \left( E_{c\mb{k}}-E_{v\mb{k}'}\right) \delta_{\mb{k},\mb{k}'} \delta_{c,c'}\delta_{v,v'} + K^x_{vc\mb{k},v'c'\mb{k}'} + K^d_{vc\mb{k},v'c'\mb{k}'}.
\end{equation}

\noi
The first term on the right-hand side of Eq. \ref{eq: H_exc_vck} is called the \textit{diagonal} term and corresponds to the IP approximation\footnote{In the \textit{diagonal} term we will consider here differences of KS energies and not of quasi-particle energies (obtained, for example, within the $GW$ approximation or from Koopmans' compliant spectral functionals~\cite{Nguyen2018}). In our implementation the KS energies can be corrected by means of an user-defined scissor operator that performs a rigid shift of the bands.}. Indeed it is worth noting that Eq. \ref{eq: eps_BSE_optic} reduces to Eq.~\ref{eq: eps_IP_inter} (in case of insulators) if the other remaining two terms in Eq.~\ref{eq: H_exc_vck}, that give rise to the electron-hole interactions, are neglected.\\
$K^d$, called \textit{direct} term, describes the attraction between the electron and hole involving the screened Coulomb interaction $W$ while $K^x$, called \textit{exchange} term, describes the repulsion between the electron and hole involving the bare Coulomb interaction $v$. Their explicit expressions are
\begin{align}
K^x_{vc\mb{k},v'c'\mb{k}'} &= \iint  d\mb{x} \, d\mb{x}' \, \psi^*_{c\mb{k}}(\mb{x}) \psi_{v\mb{k}}(\mb{x}) v(\mb{r},\mb{r}') \psi_{c'\mb{k}'}(\mb{x}') \psi^*_{v'\mb{k}'}(\mb{x}'),   \\
K^d_{vc\mb{k},v'c'\mb{k}'} &= - \iint  d\mb{x} \, d\mb{x}' \, \psi^*_{c\mb{k}}(\mb{x}) \psi_{c'\mb{k}'}(\mb{x}) W(\mb{r},\mb{r}') \psi_{v\mb{k}}(\mb{x}') \psi^*_{v'\mb{k}'}(\mb{x}').
\end{align} 

\noi 
In the non-collinear case the notation $\int d\mb{x}$ implies that we perform both an integration over the space variable $\mb{r}$ and a summation over the spin variable $\sigma$ $\left( \int  d\mb{x} \equiv \sum_{\sigma} \int d\mb{r} \right)$.

\section{Implementation} \label{section: implementation}

\subsection{Optimal basis (OB)} \label{section: implementation_ob}

\noi
The basic idea of the optimal basis (OB) method~\cite{Shirley1996} is to obtain a reduced set of basis functions, indicated with the notation $\{b_i\}$, to represent the periodic part of the Bloch wavefunctions at any k-point inside the BZ. 
The OB $\{b_i\}$ are constructed starting from the periodic KS states $\{ u_{n\mb{k}} \}$ calculated on an initial grid of $N_k$ k-points. The band index $n$, which ranges  from 1 to $N$, runs over the selected $N_v$ top valence and $N_c$ lowest conduction  bands ($N = N_v + N_c$). We use  a Gram-Schimdt orthonormalization algorithm with a threshold $s_b$  which proceeds k-point by k-point, as illustrated in Algorithm~\ref{ALGORITHM:OB}. With this approach we disregard basis vectors which would contribute marginally (i.e. below the threshold) to the periodic KS states; the dimension $N_b$ of the OB is directly dictated by the threshold $s_b$.\\ 
It is worth noting that the algorithm proceeds on blocks of $N$ states at each time. This permits, on one hand, the use of efficient BLAS matrix-matrix multiplication routines for the orthonormalization and, on the other hand, to avoid dealing with too large matrices, as it would be the case when working with all the periodic KS states at once.

\begin{algorithm}
\caption{Calculate optimal basis vectors $|b_{i}\rangle$.}
\label{ALGORITHM:OB}
\begin{algorithmic}
\State $s_b \gets$ user-defined parameter
\State $N_b \gets N$ 
\For{$i \gets 1,  N$}
\State $|b_{i}\rangle \gets |u_{i\mb{k}_1}\rangle$ 
\EndFor
\For{$l \gets 2,  N_k$}
\For{$i \gets 1,  N$}
\State $|\tilde{u}_{i\mb{k}_l}\rangle \gets |u_{i\mb{k}_l}\rangle-\sum_{j=1,N_b}|b_j\rangle\langle b_j|u_{i\mb{k}_l}\rangle$ 
\EndFor
\State $N_b' \gets N_b$
\For{$i \gets 1,  N$}
\State $|\tilde{u}'_{i\mb{k}_l}\rangle \gets |\tilde{u}_{i\mb{k}_l}\rangle-\sum_{j=N_b'+1,N_b}|b_j\rangle\langle b_j|\tilde{u}_{i\mb{k}_l}\rangle$ 
\If{$\langle\tilde{u}'_{i\mb{k}_l}|\tilde{u}'_{i\mb{k}_l}\rangle \geq s_b$}
\State $\alpha \gets \langle\tilde{u}'_{i\mb{k}_l}|\tilde{u}'_{i\mb{k}_l}\rangle$
\State $N_b \gets N_b+1$
\State $ |b_{N_b}\rangle \gets \frac{1}{\sqrt[]{\alpha}}|\tilde{u}'_{i\mb{k}_l}\rangle$
\EndIf
\EndFor
\EndFor
\end{algorithmic}
\end{algorithm}

\noindent
Using this basis we can approximate a generic, periodic KS state with wavevector $\mb{k}$ and band index $n$ as
\begin{equation} \label{eq: OB_basis}
u_{n\mb{k}}(\mb{x}) \simeq \sum_{i=1}^{N_b} \tilde{b}_i^{n\mb{k}} \, b_i(\mb{x}).
\end{equation}

\subsection{IP} \label{section: implementation_ip}
\noi 
Once the OB is constructed it is possible to obtain the periodic part of the Bloch wavefunctions at a generic k-point following the interpolation procedure described in Ref.~\cite{Prendergast2009}. This  allows to perform the fine samplings of the BZ required in Eq.~\ref{eq: eps_IP_inter} and in Eq.~\ref{eq: eps_IP_intra} for metals.
For this purpose, we need to construct the matrix elements of the k-dependent KS Hamiltonian $H^{\text{KS}}(\mb{k}) \equiv e^{-i\mb{k}\cdot \mb{r}}H^{\text{KS}} e^{i\mb{k}\cdot \mb{r}}$ in terms of the OB. By diagonalizing this matrix we can then obtain the coefficients $\tilde{b}_i^{n\mb{k}}$ and the band energies $E_{n\mb{k}}$ for each point $\mb{k}$ in the dense interpolation k-grid.

\subsubsection*{Band interpolation}
\noi
The operator $H^{\text{KS}}(\mb{k})$ is divided into three separate contributions (kinetic energy, local potential and non-local potential):
\begin{equation}
H^{\text{KS}}(\mb{k}) = 
\frac{(-i\nabla + \mb{k})^2}{2} + V^{loc} + V^{nl}(\mb{k}).
\end{equation}

\noi
In our implementation the OB functions are expanded in a plane-wave basis as
\begin{equation}
b_i(\mb{x}) = \sum_{\mb{G}} e^{i\mb{G}\cdot \mb{r}} \, b_i(\mb{G}) ,
\end{equation}

\noi
where the sum is over the reciprocal lattice vectors $\mb{G}$ and \{$b_i(\mb{G})$\} is the set of Fourier coefficients of $b_i(\mb{x})$. By using the expression above we obtain the matrix elements of the KS Hamiltonian
\begin{equation} \label{eq: H_ij_k}
H^{\text{KS}}_{ij}(\mb{k}) =  \bra{b_i }H^{\text{KS}}(\mb{k})  \ket{b_j} = 
\frac{1}{2} \big[   k^2\delta_{ij} + \mb{k}\cdot \mb{K}_{ij}^{(1)} + 
 K_{ij}^{(0)}   \big]
 +   V_{ij}^{loc}  +  V_{ij}^{nl}(\mb{k}).
\end{equation}

\noi
The terms inside the square brackets in Eq. \ref{eq: H_ij_k} refer to the matrix elements of the k-dependent kinetic energy, which have a simple quadratic polynomial dependence on $\mb{k}$. For convenience we define two additional quantities, $ \mb{K}_{ij}^{(1)}$ and $K_{ij}^{(0)}$, in terms of the Fourier coefficients $b_i(\mb{G})$:
\begin{align} \label{eq: K1_ij}
 \mb{K}_{ij}^{(1)}  &= 2 \sum_{\mb{G}}  b_i^*(\mb{G}) \,\,  \mb{G}  \,\, b_j(\mb{G}),  \\   \label{eq: K0_ij}
K_{ij}^{(0)}   &=  \sum_{\mb{G}}  b_i^*(\mb{G}) \,\,  \mb{G}^2  \,\, b_j(\mb{G}).
\end{align}

\noi
The matrix elements of the local potential do not depend on $\mb{k}$ and are easily obtained in real space using fast-Fourier transforms
\begin{equation} \label{eq: Vloc_ij}
V_{ij}^{loc} = \int d\mb{r} \,  b_i^*(\mb{x}) \,\,  V^{loc}(\mb{r})  \,\, b_j(\mb{x}).
\end{equation}

\noi
For the non-local potential, in this work we only consider norm-conserving pseudopotentials so that the non-local part of the pseudopotential can be written as
\begin{equation} \label{eq: Vnloc_ij_k}
 V_{ij}^{nl}(\mb{k}) = \sum_{\lambda} \beta_{\lambda i}^*(\mb{k}) \,\, D_{\lambda} \,\, 
\beta_{\lambda j}(\mb{k}) ,
\end{equation}

\noi
where the index $\lambda=(I,n,m,l)$ refers to the  sites $I$ of the ions in the cell together with the associated atomic quantum numbers $(n,m,l)$ and $D_{\lambda}$ are the coefficients of the pseudopotential.
In the expression above $\beta_{\lambda i}(\mb{k}) = \bra{\beta_{\lambda}}e^{i \mb{k}\cdot \mb{r}}\ket{b_i}$, where $\ket{\beta_{\lambda}}$ are the pseudopotential projector functions centered on each ionic site $I$.\\
Since $\mb{K}_{ij}^{(1)}$, $K_{ij}^{(0)}$ and $V_{ij}^{loc}$ do not depend on $\mb{k}$, they need to be calculated only once after the OB is built. The matrix elements $V_{ij}^{nl}(\mb{k})$ of the non-local part of the pseudopotential instead have a non-analytic dependence over $\mb{k}$ and thus they should be calculated for every $\mb{k}$. 
The calculation of the matrix elements $ H^{\text{KS}}_{ij}(\mb{k})$ and the subsequent diagonalization of the matrix for each point $\mb{k}$ give the coefficients $\tilde{b}_i^{n\mb{k}}$ and the band energies $E_{n\mb{k}}$ for all the bands $n \leq N_b$.

\subsubsection*{Dielectric function}
\noi
Once the coefficients and the band energies at all the needed k-points are known, we have almost all the elements necessary to calculate the IP dielectric function through Eq.~\ref{eq: eps_IP_inter} and Eq.~\ref{eq: eps_IP_intra}. For this purpose we also need the matrix elements of the k-dependent velocity operator
\begin{equation} \label{eq: velocity_k}
\mb{v}(\mb{k}) = -i [\mb{r}, H^{\text{KS}}(\mb{k})] =  -i \nabla + \mb{k} - i [\mb{r}, V^{nl}(\mb{k})].
\end{equation}

\noi
In terms of the OB these are given by
\begin{equation}
\begin{split} \label{eq: velocity_nn'_k}
\bra{u_{n\mb{k}}}  \mb{v}(\mb{k})  \ket{u_{n'\mb{k}}} & = 
 \frac{1}{2} \sum_{i,j=1}^{N_b} (\tilde{b}_i^{n\mb{k}})^*\,\tilde{b}_j^{n'\mb{k}} \, \mb{K}_{ij}^{(1)}  + \mb{k} \braket{u_{n\mb{k}}   |u_{n'\mb{k}} } + \\
& + (-i) \sum_{i,j=1}^{N_b}  (\tilde{b}_i^{n\mb{k}})^*\, \tilde{b}_j^{n'\mb{k}} \,     \bra{b_i}  [\mb{r}, V^{nl}(\mb{k})]   \ket{b_j}.
\end{split}
\end{equation}

\noi
The local contribution to the velocity matrix elements (first two terms in the right-hand side of Eq.~\ref{eq: velocity_nn'_k}) are easily obtained once the KS Hamiltonian matrix in Eq.~\ref{eq: H_ij_k} is diagonalized.
On the other hand, the matrix elements $\bra{b_i}  [\mb{r}, V^{nl}(\mb{k})]   \ket{b_j}$ of the commutator of the non-local part of the pseudopotential need to be computed and, having a non-analytic dependence over $\mb{k}$, they should be calculated for each $\mb{k}$.

\subsection{BSE} \label{section: implementation_bse}
\noi
For BSE calculations in finite-gap systems we do not interpolate over the k-points, as in the IP calculation, but we work with the initial k-grid as there is no need to perform an interpolation since significantly coarser k-grids with respect to metals are usually sufficient to obtain converged results. In this case the band energies are known from the initial DFT calculation, while the coefficients $\tilde{b}_i^{n\mb{k}}$ are simply calculated through the scalar products
\begin{equation}
\tilde{b}_i^{n\mb{k}} = \braket{b_i|u_{n\mb{k}}}.
\end{equation}


\subsubsection*{Optimal product basis (OPB)}

\noi
We then build a basis, that we call optimal product basis (OPB), for representing in real space products of the periodic parts of the Bloch wavefunctions at the same k-point: $u_{n\mathbf{k}}^{*}\left(\mathbf{x}\right)u_{n'\mathbf{k}}\left(\mathbf{x}\right)$.
We construct this optimally small basis $\left\{ B_{\alpha}\right\} $ using a Gram-Schmidt orthonormalisation algorithm with a threshold  starting from the products $b_{i}^{*}\left(\mathbf{x}\right)b_{j}\left(\mathbf{x}\right)$ of OB functions obtained in the first step.
The threshold $s_{p}$ controls the precision of the OPB and hence its dimension, that we indicate with $N_{p}$. We outline this strategy in Algorithm~\ref{ALGORITHM:OPB}.

\begin{algorithm}
\caption{Calculate optimal product basis vectors $|B_{\alpha}\rangle$.}
\label{ALGORITHM:OPB}
\begin{algorithmic}
\State $s_p \gets$ user-defined parameter
\State $N_p \gets 0$
\For{$i \gets 1,  N_b$}
\For{$j \gets 1,  N_b$}
\State $\tilde{B}_j(\mathbf{x})\gets b_i^*(\mathbf{x})b_j(\mathbf{x})$
\EndFor
\For{$j \gets 1,  N_b$}
\State $|\tilde{B}'_j\rangle\gets |\tilde{B}_j\rangle -\sum_{\alpha=1,N_p}|B_\alpha\rangle\langle B_\alpha|\tilde{B}_j\rangle  $
\EndFor
\State $N'_p \gets 0$
\For{$j \gets 1,  N_b$}
\State $|\tilde{B}''_j\rangle\gets |\tilde{B}'_j\rangle -\sum_{\alpha=N_p+1,N_p+N_p'}|B_\alpha\rangle\langle B_\alpha|\tilde{B}'_j\rangle  $
\If{$\langle\tilde{B}''_j|\tilde{B}''_j\rangle \geq s_p$}
\State $N'_p\gets N'_p+1$
\State $|B_{N'_p}\rangle\gets \frac{1}{\sqrt[]{\langle\tilde{B}''_j|\tilde{B}''_j\rangle}}|\tilde{B}''_j\rangle$
\EndIf
\EndFor
\State $N_p\gets N_p+N'_p$
\EndFor
\end{algorithmic}
\end{algorithm}
\noi
Also in this case we can take advantage of the BLAS library for matrix-matrix multiplications, avoiding at the same time to deal with large matrices. 
Using this new basis, we approximate the product of two OB functions as
\begin{equation}
b_{i}^{*}(\mathbf{x}) \, b_{j}(\mathbf{x}) \simeq  \sum_{\alpha=1}^{N_p}F_{ij}^{\alpha} \, B_{\alpha}(\mathbf{x}),
\end{equation}
with
\begin{equation}
F_{ij}^{\alpha}=\braket{B_{\alpha}| b_{i}^{*}b_{j}}.
\end{equation}

\noi
We can now express products of periodic KS states as
\begin{equation}
\ket{u_{n\mathbf{k}}^*u_{n'\mathbf{k}}} 
\simeq \sum_{i,j=1}^{N_b} (\tilde{b}_i^{n\mb{k}})^*\,\tilde{b}_j^{n'\mb{k}} \ket{b_i^* b_j} 
\simeq \sum_{\alpha=1}^{N_p} \sum_{i,j=1}^{N_b}  (\tilde{b}_i^{n\mb{k}})^*\,\tilde{b}_j^{n'\mb{k}} F_{ij}^{\alpha} \ket{B_{\alpha}}.
\end{equation}

\noi
We then define
\begin{equation}
J_{\alpha,nn'}^{\mathbf{k}}=  
\sum_{i,j=1}^{N_b} (\tilde{b}_i^{n\mb{k}})^*\,\tilde{b}_j^{n'\mb{k}}F_{ij}^{\alpha} ,
\end{equation}
\noi
so that, similarly to Eq. \ref{eq: OB_basis}, $u_{n\mathbf{k}}^{*}\left(\mathbf{x}\right)u_{n'\mathbf{k}}\left(\mathbf{x}\right)$ is approximated as
\begin{equation}
\ket{u_{n\mathbf{k}}^{*}u_{n'\mathbf{k}} } \simeq \sum_{\alpha=1}^{N_p}J_{\alpha,nn'}^{\mathbf{k}}\ket{B_{\alpha}} . \label{eq:uuj}
\end{equation}

\subsubsection*{Exchange and direct term}
\noi
The reduction of the cost of the BSE calculation is achieved by evaluating only on the product basis $\left\{ B_{\alpha}\right\}$ the matrix elements of the bare ($v$) and of the screened ($W$) Coulomb interactions required for the exchange and the direct terms of the excitonic Hamiltonian. The number of such elements is much smaller then all the  elements defining the $K^x_{vc\mb{k},v'c'\mb{k}'}$ and $K^d_{vc\mb{k},v'c'\mb{k}'}$ matrices.  These are 
then  obtained through:
\begin{eqnarray*}
K^x_{vc\mb{k},v'c'\mb{k}'} & = & \int d\mathbf{x}\, d\mathbf{x'} \, u_{\mathbf{k}c}^{*}\left(\mathbf{x}\right)u_{\mathbf{k}v}\left(\mathbf{x}\right)v\left(\mathbf{r},\mathbf{r}'\right)u_{\mathbf{k'}c'}\left(\mathbf{x}'\right)u_{\mathbf{k'}v'}^{*}\left(\mathbf{x}'\right)\\
 & \simeq & \sum_{\alpha,\beta=1}^{N_p} \left( J_{\alpha,vc}^{\mathbf{k}} \right)^* J_{\beta,v'c'}^{\mathbf{k'}}\int d\mathbf{x}\, d\mathbf{x'} \,B_{\alpha}^{*}\left(\mathbf{x}\right)v\left(\mathbf{r},\mathbf{r}'\right)B_{\beta}\left(\mathbf{x}'\right)\\
 & = & \sum_{\alpha,\beta=1}^{N_p} \left( J_{\alpha,vc}^{\mathbf{k}} \right)^*J_{\beta,v'c'}^{\mathbf{k'}}V_{\alpha\beta}^{\mb{0}} ,
\end{eqnarray*}

\noi
where we have defined
\begin{equation} \label{eq: V_ab_k''}
V_{\alpha\beta}^{\mathbf{k}''}=\int d\mathbf{x}\,d\mathbf{x'}\,B_{\alpha}^{*}\left(\mathbf{x}\right)
e^{i\mathbf{k}''\cdot\mathbf{r}}v\left(\mathbf{r},\mathbf{r}'\right)e^{-i\mathbf{k}''\cdot\mathbf{r}'} B_{\beta}\left(\mathbf{x}'\right).
\end{equation}

\noi
We divide the screened Coulomb interaction in a bare and a correlation part: $W=v+W_c$. The bare part of the direct operator reads
\begin{eqnarray*}
K_{vc\mb{k},v'c'\mb{k}'}^{d,bare} & = & -\int d\mathbf{x}\,d\mathbf{x'}\,u_{c\mathbf{k}}^{*}\left(\mathbf{x}\right)u_{c'\mathbf{k'}}\left(\mathbf{x}\right)e^{i\left(\mathbf{k}'-\mathbf{k}\right)\cdot\mathbf{r}}v\left(\mathbf{r},\mathbf{r}'\right)e^{-i\left(\mathbf{k}'-\mathbf{k}\right)\cdot\mathbf{r}}u_{v\mathbf{k}}\left(\mathbf{x}'\right)u_{v'\mathbf{k'}}^{*}\left(\mathbf{x}'\right)\label{eq:kdb}\\
 & \simeq & -\sum_{\alpha,\beta=1}^{N_p} \left(G_{\alpha,c'c}^{\mathbf{k}'\mathbf{k}}\right)^* G_{\beta,v'v}^{\mathbf{\mathbf{k}'}\mathbf{k}}\int d\mathbf{x}\,d\mathbf{x'}\,B_{\alpha}^{*}\left(\mathbf{x}\right)
e^{i\left(\mathbf{k}'-\mathbf{k}\right)\cdot\mathbf{r}}v\left(\mathbf{r},\mathbf{r}'\right)e^{-i\left(\mathbf{k}'-\mathbf{k}\right)\cdot\mathbf{r}'}
B_{\beta}\left(\mathbf{x}'\right)\nonumber \\
 & = & -\sum_{\alpha,\beta=1}^{N_p}\left(G_{\alpha,c'c}^{\mathbf{k}'\mathbf{k}}\right)^*
 G_{\beta,v'v}^{\mathbf{\mathbf{k}'}
 \mathbf{k}}V_{\alpha\beta}^{\mathbf{\mathbf{k}'}-\mathbf{k}} , \nonumber 
\end{eqnarray*}
where we have introduced the matrices
\begin{equation}
G_{\alpha,nn'}^{\mathbf{k}\mathbf{k}'}=\sum_{i,j=1}^{N_b}(\tilde{b}_i^{n\mb{k}})^*\,\tilde{b}_j^{n'\mb{k}'} F_{ij}^{\alpha}.
\end{equation}

\noi
The correlation part of the direct operator instead reads
\begin{eqnarray*}
K_{vc\mb{k},v'c'\mb{k}'}^{d,c} & = & -\int d\mathbf{x}\,d\mathbf{x'}\,u_{c\mathbf{k}}^{*}\left(\mathbf{x}\right)u_{c'\mathbf{k'}}
\left(\mathbf{x}\right)e^{i\left(\mathbf{k}'-\mathbf{k}\right)\cdot\mathbf{r}}W_{c}\left(\mathbf{r},\mathbf{r}'\right)e^{-i\left(\mathbf{k}'-\mathbf{k}\right)\cdot\mathbf{r}'}u_{v\mathbf{k}}\left(\mathbf{x}'\right)u_{v'\mathbf{k'}}^{*}\left(\mathbf{x}'\right)\label{eq:kdc}\\
 & \simeq & -\sum_{\alpha,\beta=1}^{N_{p}}\left( G_{\alpha,c'c}^{\mathbf{k}',\mathbf{k}}\right)^* G_{\beta,v'v}^{\mathbf{\mathbf{k}'}\mathbf{k}}\int d\mathbf{x}\,d\mathbf{x'}\,e^{i\left(\mathbf{k}'-\mathbf{k}\right)\cdot\mathbf{r}}
 B_{\alpha}^{*}\left(\mathbf{x}\right)W_c\left(\mathbf{r},\mathbf{r}'\right)e^{-i\left(\mathbf{k}'-\mathbf{k}\right)\cdot\mathbf{r}'}B_{\beta}\left(\mathbf{x}'\right)\nonumber \\
 & = & -\sum_{\alpha,\beta=1}^{N_p}\left( G_{\alpha,c'c}^{\mathbf{k}'\mathbf{k}}\right)^*\,
 G_{\beta,v'v}^{\mathbf{\mathbf{k}'}\mathbf{k}}
 W_{c,\alpha\beta}^{\mathbf{\mathbf{k}'}-\mathbf{k}} , \nonumber  
\end{eqnarray*}

\noi
where we have defined
\begin{equation} \label{eq: W_ab_k''}
W_{c,\alpha\beta}^{\mathbf{k}''}=\int d\mathbf{x}\, d\mathbf{x'}\,B_{\alpha}^{*}\left(\mathbf{x}\right)
e^{i\mathbf{k}''\cdot\mathbf{r}}W_{c}\left(\mathbf{r},\mathbf{r}'\right)e^{-i\mathbf{k}''\cdot\mathbf{r}'}B_{\beta}\left(\mathbf{x}'\right).
\end{equation}

\subsubsection*{Treatment of the Coulomb interaction}
\noi
We evaluate Eq.~\ref{eq: V_ab_k''} and Eq.~\ref{eq: W_ab_k''} in a plane-wave basis where the Coulomb interaction is diagonal. However, some care is needed because in reciprocal space terms like 
\begin{equation}
\frac{1}{\left|\mathbf{k}' - \mathbf{k} +\mathbf{G}\right|^{2}} , 
\end{equation}
appear, which diverge for $\mathbf{k}' - \mathbf{k}\rightarrow0,\mathbf{\;G}\rightarrow0$ (it is worth noting that these divergencies are not physical but come from the numerical discretization of the BZ). To avoid these divergencies we perform the following replacement
\begin{equation}
\frac{1}{\left|\mb{k}' - \mb{k} + \mathbf{G}\right|^{2}}=\frac{\Omega N_{q}}{\left(2\pi\right)^{3}}\int_{\text{BZR}}d\mathbf{q'}\frac{1}{\left|\mb{k}' - \mb{k}+\mathbf{q'}+\mathbf{G}\right|^{2}},
\end{equation}
i.e. we integrate numerically over the reduced volume of the Brillouin's zone (BZR) centered in $\mb{k}' - \mb{k}$ and of volume $\frac{\left(2\pi\right)^{3}}{\Omega N_{q}}$ , where $\Omega$ is the volume of the simulation cell and $N_{q}$ is the total number of k-points of a uniform mesh in the BZ. The integration is performed as a sum over an equally-spaced grid which does not contain the $\Gamma$ point. This is equivalent to the strategy of Ref.~\cite{RIM_method}, where instead the sum is performed through Monte-Carlo sampling.
In the present implementation the correlation part of the screened Coulomb interaction ($W_c$) is either approximated using the model described in Ref.~\cite{Bechstedt1992} or is taken from a $GW$ calculation performed with the \texttt{GWL} package of the QE distribution~\cite{Giannozzi2017}. \\ 
The model screened Coulomb interaction is obtained from the following expression for the inverse microscopic dielectric function
\begin{equation}
\label{EQ:model_eps}
\varepsilon^{-1}_{\text{model}}(\mathbf{G},\mathbf{G}';\mathbf{q} ) = \delta(\mathbf{G},\mathbf{G}')\left[ 1 - \left(1-\epsilon_m^{-1}\right)e^{\frac{-2\pi|\mathbf{q}+\mathbf{G}|^2}{4\overline{\lambda}^2}}\right].
\end{equation}
\\
The model depends on two parameters: the macroscopic dielectric constant $\epsilon_m$ and the screening length $\overline{\lambda}$. From $\varepsilon^{-1}_{\mathrm{model}}(\mathbf{G},\mathbf{G}';\mathbf{q} )$ an approximated screened interaction is earned
\begin{equation}
    W_{\mathrm{model}}(\mathbf{G},\mathbf{G}';\mathbf{q} ) =\varepsilon^{-1}_{\text{model}}(\mathbf{G},\mathbf{G}';\mathbf{q} )
\frac{1}{\left| \mathbf{G}+\mathbf{q}\right|^2 }
\end{equation} 
where $\mathbf{q}=\mathbf{k}'-\mathbf{k}$ is the transferred momentum. $W_{\mathrm{model}}(\mathbf{G},\mathbf{G}';\mathbf{q})$ is then Fourier transformed to be inserted in Eq.~\ref{eq: W_ab_k''}. 

\noi
If $W_c$ is taken instead from a \texttt{GWL} calculation, some care is required because \texttt{GWL} samples the BZ at the $\Gamma$ point only, even if the long-range parts of the dielectric matrix (commonly known as ``wings" of the matrix) are properly calculated on a discrete k-points mesh.  In this way the convergence of $GW$ energies is achieved for not too large simulation cells. Indeed, the \texttt{GWL} code is optimized to work with large model structures, as described in Ref.~\cite{Umari2010}. However, denser k-point samplings are required for a subsequent BSE calculation. For this purpose, we extrapolate $W_c$ at any k-point $\mathbf{q}$ from the polarizability matrix $\Pi$ calculated at the $\Gamma$ point and at zero frequency
\begin{equation}
W_{c}(\mathbf{G},\mathbf{G}';\mathbf{q} ) = \sum_{\mu\nu}
\frac{\Phi_{\mu}\left(\mathbf{G}\right)}{\left|\mathbf{q}+\mathbf{G}\right|^{2}}\Pi_{\mu\nu}\frac{\Phi_{\nu}\left(\mathbf{G'}\right)}{\left|\mathbf{q}+\mathbf{G}'\right|^{2}}  \label{eq: W_c_GWW}
\end{equation}

\noi
where $\left\{ \Phi_{\mu}\right\} $ is a generic and  real basis set used to represent the polarizability operator. Although in our calculations the $\left\{ \Phi_{\mu}\right\} $ are found according to the so-called optimal polarizability basis recipe \cite{Umari09,Umari2010}, our formalism remains unaltered if a plane-waves basis set is chosen.
We stress that the expression above is formally exact only at $\mb{q}=\mb{0}$ and its use at a generic $\mb{q}=\mb{k}'-\mb{k} \neq \mb{0}$ is  justified only if we assume that KS bands and wavefunctions do not change significantly at different k-points. This flat band approximation works well in the case of large simulation cells (or supercells).

%
\subsubsection*{Single excitonic states}
\noi
We find the lowest eigenvalues and eigenvectors of the excitonic Hamiltonian matrix by solving Eq.~\ref{EQ:excitonic_hamiltonian}. Instead of the direct diagonalization of the entire matrix, which would be too costly, we use an iterative conjugate-gradient (CG) algorithm~\cite{Numerical_recipies} so that only the lowest states are found. 
The CG algorithm proceeds optimizing the search direction: at each minimization step the minimum along the search direction is found, assuming  a parabolic behaviour for the excitonic energy. The minimum is found from the energy and its line derivative at the initial point and the energy  calculated at a trial point, which is defined by a factor $\mu$ and is given as an input parameter.

\subsubsection*{Dielectric function}
\noi
For the calculation of the dielectric function given in Eq. \ref{eq: eps_BSE_optic} we  neglect, in the present implementation, the contribution coming from the non-local commutator (see Eq.~\ref{eq: velocity_k}). 
In principle, all the excitonic  eigenvectors are required so that  a full diagonalization of the excitonic Hamiltonian would be necessary. However, iterative algorithms can be used for evaluating directly the frequency-dependent dielectric function. We adopt here the Haydock's recursive method~\cite{Benedict1999}, so that we can calculate $\varepsilon_{\text{BSE}}(\hat{\mb{q}},\omega)$ without the need to diagonalize the excitonic Hamiltonian. Within this formalism Eq.~\ref{eq: eps_BSE_optic} can be rewritten as
\begin{equation}
\label{EQ:eps_BSE}
\varepsilon_{\mathrm{BSE}}(\hat{\mb{q}},\omega)=1-\frac{4\pi}{V}\sum_{\lambda}\sum_{vc\mb{k}}\frac{P^*_{vc\mb{k}}A_{\lambda}^{(vc\mb{k})*}A_{\lambda}^{(vc\mb{k})}{P_{vc\mb{k}}}}{\omega-\tilde{E}_\lambda+i\eta}.
\end{equation}
\noindent
In the case of isolated systems the components of the vectors $|P\rangle$ are 
\begin{equation}
P_{vc\mb{k}}=\langle\psi_{v\mathbf{k}}|\hat{\mb{q}} \cdot \mathbf{r}|\psi_{c\mathbf{k}}\rangle
\end{equation}
\noindent
for which only $\Gamma$-point sampling is meaningful, while for extended systems they read
\begin{equation}
P_{vc\mb{k}}=\frac{\bra{\psi_{v\mb{k}}} \hat{\mb{q}} \cdot \mathbf{v} \ket{\psi_{c\mb{k}}}}{E_{c\mb{k}}-E_{v\mb{k}}}.
\end{equation}
This allows to eliminate the explicit sum over $\lambda$ in Eq.~\ref{EQ:eps_BSE}:
\begin{equation}
\varepsilon_{\text{BSE}}(\hat{\mb{q}},\omega)=1 - \frac{4\pi}{V}
\bra{P}\left(\omega-H^{exc}+i\eta\right)^{-1}\ket{P}.
\end{equation}
The Haydock's recursive method provides us with two series of coefficients $\alpha_i$ and $\beta_i$ so that we can write the continuous fraction as
\begin{equation}
\label{EQ:continuos_fraction}
\bra{P}\left(\omega-H^{exc}+i\eta\right)^{-1}\ket{P}=
\frac{1}{\omega+i\eta-\alpha_1-\frac{\beta_1^2}{\omega+i\eta-\alpha_2-\frac{\beta_2^2}{\omega+i\eta-\alpha_3\dots}}}.
\end{equation}
The factors $\alpha_i$ and $\beta_i$ converge to the values $\alpha_\infty$ and $\beta_\infty$ for large $i$. This permits to terminate the fraction substituting the final $\beta_n^2/\cdot$ with the terminator $\beta_n^2/T(\omega)$, where
\begin{equation}
T(\omega)=\frac{\omega-a_\infty+\sqrt{(a_\infty-\omega)^2}-4b_\infty^2}{2}.
\end{equation}

\section{Details of the code}

\begin{figure}[H]
\centering
\includegraphics[width=\textwidth]{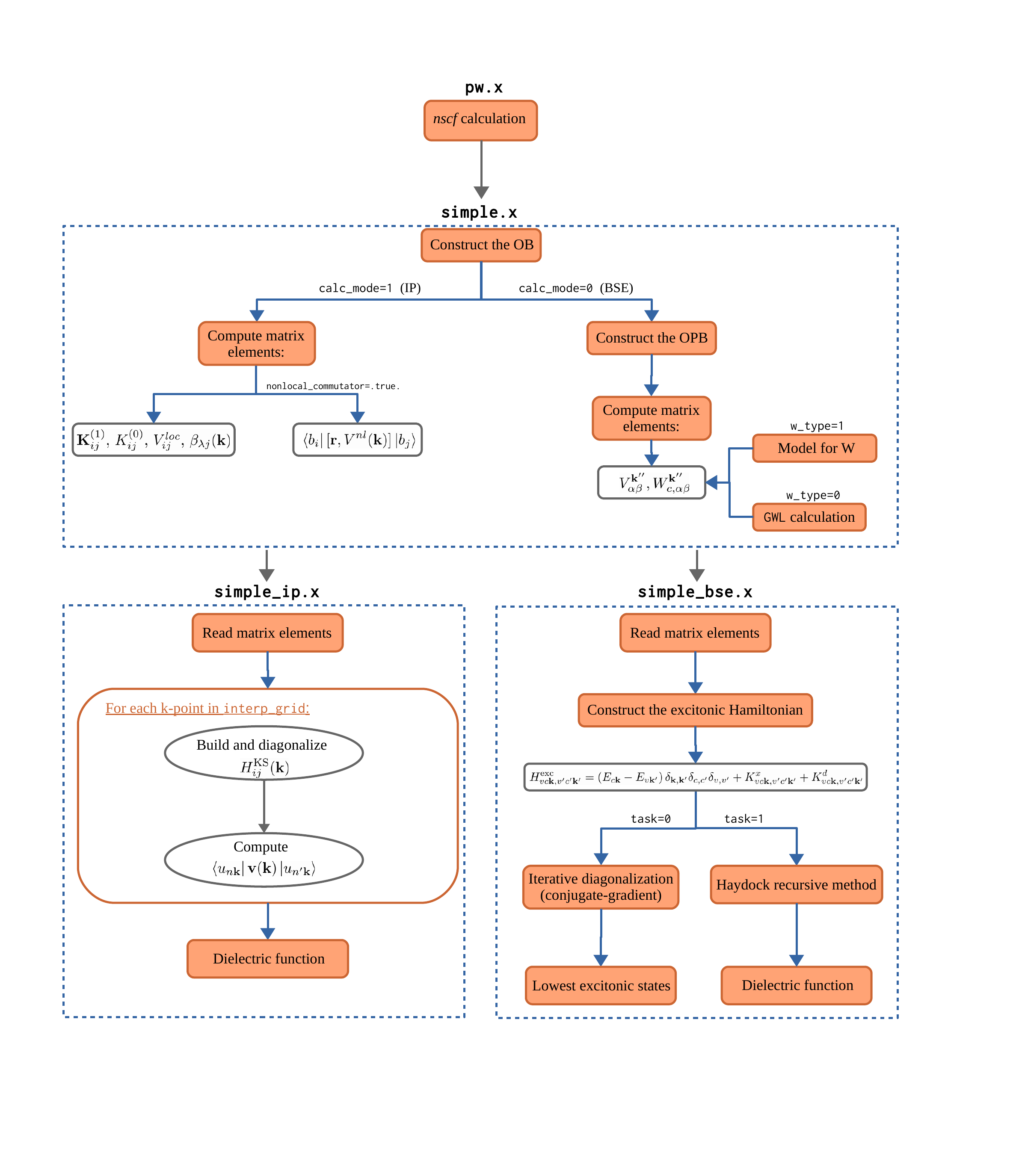}
\caption{Schematic structure of the \texttt{SIMPLE} code. Two steps are required for the calculation of the optical properties. In the first step \texttt{simple.x} reads the data produced by a previous \textit{nscf} calculation performed with \texttt{pw.x} of QE, builds the optimal basis and saves to disk the matrix elements needed for the subsequent IP or BSE calculation. In the second step, \texttt{simple\_ip.x} or \texttt{simple\_bse.x} reads the matrix elements and performs the actual IP or BSE calculation, respectively.}
\label{fig: code_structure}
\end{figure}

\noi
The \texttt{SIMPLE} code is implemented as a post-processing code of the QE distribution, directly inserted within the \texttt{GWL} package, and its structure is schematically depicted in Fig.~\ref{fig: code_structure}. \\
It is divided into three different executable: \texttt{simple.x}, \texttt{simple\_ip.x} and \texttt{simple\_bse.x}. These could be easily interfaced with different DFT codes as they are separate from the other routines of \texttt{GWL}. The executable \texttt{simple.x} builds the OB and saves to disk all the relevant matrix elements needed for the calculation of the optical properties. The actual IP or BSE calculation is then performed by \texttt{simple\_ip.x} or \texttt{simple\_bse.x}, respectively.
\texttt{simple.x} uses the QE environment while \texttt{simple\_ip.x} and \texttt{simple\_bse.x} are independent from this environment but make use of several routines and modules of QE. Throughout the \texttt{SIMPLE} code, MPI parallelization is exploited and linear algebra operations are efficiently performed through calls to BLAS and LAPACK libraries. 
The code works only with norm-conserving pseudopotentials. Spin-orbit interactions are implemented and can be included for both IP and BSE calculations. \\
Two test examples, i.e. bulk silver for \texttt{simple\_ip.x} and bulk silicon for \texttt{simple\_bse.x}, are included within the \texttt{SIMPLE} code. 

\subsubsection*{simple.x}
\noi
\texttt{simple.x} relies, as a starting point, on the results of a \textit{nscf} calculation of QE performed on a uniform grid of k-points without the use of symmetry (\texttt{nosym=.true.} and \texttt{noinv=.true.} in the input of \texttt{pw.x}). A list of the input parameters of \texttt{simple.x} is reported in Table~\ref{tab: input_simple}.\\
As a first step it builds the OB functions as described in Section \ref{section: implementation_ob} where the threshold $s_b$ controlling the precision of the OB is specified by the input variable \texttt{s\_bands} (in bohr$^3$). The number of valence bands $N_v$ included, starting from the highest occupied band (HOMO), and the number of conduction bands $N_c$ included, starting from the lowest unoccupied band (LUMO), are specified by \texttt{num\_val} and \texttt{num\_cond}, respectively. 
It then computes and saves to disk the matrix elements needed for the subsequent BSE calculation (if \texttt{calc\_mode=0} in input) or for the subsequent IP calculation (if \texttt{calc\_mode=1} in input).\\
For a IP calculation the k-independent matrix elements $\mb{K}_{ij}^{(1)}$, $K_{ij}^{(0)}$ and $V_{ij}^{loc}$ (see Eq.~\ref{eq: K1_ij}, Eq.~\ref{eq: K0_ij} and Eq.~\ref{eq: Vloc_ij} respectively) of the KS Hamiltonian in the OB are computed. For the non-local part of the pseudopotential we store the k-dependent projectors $\beta_{\lambda j}(\mb{k})$ of Eq.~\ref{eq: Vnloc_ij_k}, as originally proposed in Ref.~\cite{Prendergast2009}, at each k-point $\mb{k}$ on a uniform $n_1\times n_2\times n_3$ k-grid in the BZ defined by the input parameter \texttt{nkpoints}.
Besides, if \texttt{nonlocal\_commutator=.true.} as it is by default, the matrix elements $\bra{b_i}  [\mb{r}, V^{nl}(\mb{k})]   \ket{b_j}$ giving the non-local contribution to the velocity operator (see last term of Eq.~\ref{eq: velocity_nn'_k}) are also computed on the k-grid defined by \texttt{nkpoints}. \\ 
The calculation of the non-local commutator is the most time and memory consuming part of the simulation. 
By neglecting this term the computational cost is significantly reduced but the error introduced in the final optical spectra should be checked for each system and pseudopotential considered.\\
\noi
For a BSE calculation the OPB is first constructed following the procedure described in Section \ref{section: implementation_bse} where the threshold $s_p$ controlling the precision of the OPB is specified by the input parameter \texttt{s\_product} (in bohr$^3$). 
The matrix elements in the OPB defined in Eq. \ref{eq: V_ab_k''} and Eq. \ref{eq: W_ab_k''} are then computed ($\mb{k}''$ in Eq. \ref{eq: V_ab_k''} and Eq. \ref{eq: W_ab_k''} is constructed from the same input parameter \texttt{nkpoints} described above which now, differently from the IP case, has to be the same as the initial \textit{nscf} k-grid). \\
The screened Coulomb interaction is either approximated with the model of Eq.~\ref{EQ:model_eps}  if \texttt{w\_type} is set to 1, or is taken from a previous \texttt{GWL} calculation if \texttt{w\_type} is set to 0.
In the first case, the macroscopic dielectric constant (\texttt{epsm}) and the screening length (\texttt{lambdam}) must be given. In the second case, a complete $G_0W_0$ calculation  is performed in order to obtain the screened Coulomb interaction. The polarizability is represented either on a common plane-waves basis or on an optimal polarizability basis set. In both cases its  dimension \texttt{numpw} is required as an input parameter. If \texttt{l\_truncated\_coulomb} is \texttt{.false.} the complete long-range Coulomb interaction  is considered, in the opposite case the Coulomb interaction is truncated at a distance \texttt{truncation\_radius}. The latter case is meaningful exclusively for $\Gamma$-only sampling.

\subsubsection*{simple\_ip.x}
\noi 
The IP calculation is performed by the executable \texttt{simple\_ip.x} and a list of its input parameters is reported in Table~\ref{tab: input_simpleip}.
For a IP calculation in which a k-space interpolation is performed, the uniform k-grid specified in the starting \textit{nscf} calculation should include also the seven periodic images of $\Gamma$ at the corners of the unit cube (in units of the reciprocal lattice vectors) in order to better preserve the periodicity in reciprocal space of the k-dependent Hamiltonian (see Ref.~\cite{Prendergast2009} for a more detailed discussion). 
For this purpose we supply a \texttt{python} script (\texttt{gen\_kgrid.py}) that can generate a uniform grid of k-points for the QE input with or without periodic images of $\Gamma$ starting from a user-defined k-grid.\\
The uniform $m_1\times m_2\times m_3$ k-grid on which the IP dielectric function is computed is defined by \texttt{interp\_grid} and it should be the same as the variable \texttt{nkpoints} of \texttt{simple.x}. However, in order to reduce the computational cost, a simple linear interpolation of the non-local contributions, i.e. $\beta_{\lambda j}(\mb{k})$ and $\bra{b_i}  [\mb{r}, V^{nl}(\mb{k})]   \ket{b_j}$, is also implemented so that it is possible to use a denser k-grid in \texttt{interp\_grid} with respect to \texttt{nkpoints} (by specifying \texttt{nonlocal\_interpolation = .true.}). To have good precision it is not suggested to use a grid larger than the double of the grid specified by \texttt{nkpoints} and the reliability of the results should be always carefully checked. \\
The code performs the band interpolation on \texttt{interp\_grid} and calculates the Fermi energy if it is not already specified in input by \texttt{fermi\_energy}. The broadening type and the value of the broadening (in Ry) can be specified in input by the parameters \texttt{fermi\_ngauss} and \texttt{fermi\_degauss}, respectively.\\
The occupations of the states are then computed according to the Fermi-Dirac distribution with a broadening specified by \texttt{elec\_temp} (in Ry) that by default is set to room temperature. The IP dielectric function is then calculated  on an equispaced energy grid of \texttt{nw} points in the energy interval [\texttt{wmin}, \texttt{wmax}] (in Ry).\\
The derivative of the Fermi-Dirac distribution appearing in the expression of the Drude plasma frequency (see Eq.~\ref{eq: drude_plasma_freq}) should be in principle computed with the broadening specified by \texttt{elec\_temp}, which however can require extremely dense k-grids. Therefore, a Gaussian broadening is used instead, defined in input by \texttt{drude\_degauss}, because converged results can be obtained in practice using coarser k-grids and larger broadenings. 
The empirical interband and intraband broadenings, i.e. $\eta$ of Eq.~\ref{eq: eps_IP_inter} and $\gamma$ of Eq.~\ref{eq: eps_IP_intra}, are defined by \texttt{inter\_broadening} and \texttt{intra\_broadening}, respectively. \\
The code writes in standard output the Drude plasma frequency $\omega_{\text{D}}(\hat{\mb{q}})$, computed in the limit $\mb{q} \to \mb{0}$ along the three Cartesian directions (i.e. with $\hat{\mb{q}}=\hat{\mb{x}},\hat{\mb{y}},\hat{\mb{z}}$), and provides two data files with the real and imaginary part of the interband contribution to the IP dielectric function, $\Re{\left[\varepsilon_{\text{IP}}^{\text{inter}}(\hat{\mb{q}},\omega)\right]}$ and $\Im{\left[\varepsilon_{\text{IP}}^{\text{inter}}(\hat{\mb{q}},\omega)\right]}$, also computed along the three Cartesian limits. The real and imaginary part of the total IP dielectric function of Eq. \ref{eq: eps_IP_inter+intra} and the electron energy loss spectrum (EELS), which is given by $-\Im{\left[\varepsilon_{\text{IP}}^{-1}(\hat{\mb{q}},\omega)\right]}$, are also provided.\\
For completeness the code also computes and writes on file the total dielectric function averaged over the three Cartesian directions
\begin{equation}
\left<\varepsilon_{\text{IP}}(\omega)\right> = \frac{ \varepsilon_{\text{IP}}(\hat{\mb{x}},\omega) + \varepsilon_{\text{IP}}(\hat{\mb{y}},\omega) + \varepsilon_{\text{IP}}(\hat{\mb{z}},\omega) }{3},
\end{equation}

\noi
together with the averaged refractive index $n(\omega)$, extinction coefficient $k(\omega)$ and reflectivity at normal incidence $R(\omega)$. 
The density of states (DOS) and the joint density of states (JDOS) are also computed.

\subsubsection*{simple\_bse.x}
\noi
The BSE calculation is performed by the executable \texttt{simple\_bse.x} and a list of its input parameters is reported in Table \ref{tab: input_simplebse}. The number of terms in the excitonic Hamiltonian can be chosen through the parameter \texttt{h\_level}. Considering only the diagonal term in Eq.~\ref{eq: H_exc_vck} corresponds to  the IP approximation, adding also the exchange term gives the time-dependent Hartree approximation while adding also the bare Coulomb part of the direct term yields the time-dependent Hartree-Fock approximation. If  the correlation part of the direct term is also included, the full BSE approach is recovered.
The code supports either spin collinear calculations with doubly occupied DFT states or the general case of non collinear spin (option \texttt{spin\_state}). In the first case, exciton states have either singlet or triplet symmetry \cite{Rohlfing2000}. 
The code performs two different tasks (option \texttt{task}): the evaluation of the lowest eigenvectors of the excitonic Hamitonian and of the corresponding energies through the CG algorithm (\texttt{task}=0), and the computation of the $\alpha_i$ and $\beta_i$ coefficients of Eq.~\ref{EQ:continuos_fraction} through the Haydock's recursive scheme (\texttt{task}=1). In the latter case, three series of  coefficients relative to the three Cartesian directions are written on disk in a file named \texttt{ab\_coeff.dat} for later calculation of the continuous fraction in order to get the components 
$\varepsilon_{\text{BSE}}(\hat{\mb{x}},\omega)$, $\varepsilon_{\text{BSE}}(\hat{\mb{y}},\omega)$ and $\varepsilon_{\text{BSE}}(\hat{\mb{z}},\omega)$ 
of the complex frequency-dependent dielectric function. For this purpose we include in the distribution a basic post-processing program (\texttt{abcoeff\_to\_eps.f90}).

\section{Convergence and verification}
\noi
In this section, first we study the convergence of the optical properties calculated with the \texttt{SIMPLE} code with respect to the relevant computational parameters controlling the precision of the simulations. Subsequently we verify the code by comparing our results for test systems with other well-established and publicly available codes and/or with results in the literature.
\subsection{IP}
\begin{figure}[hbtp]
\centering
\includegraphics[width=9cm]{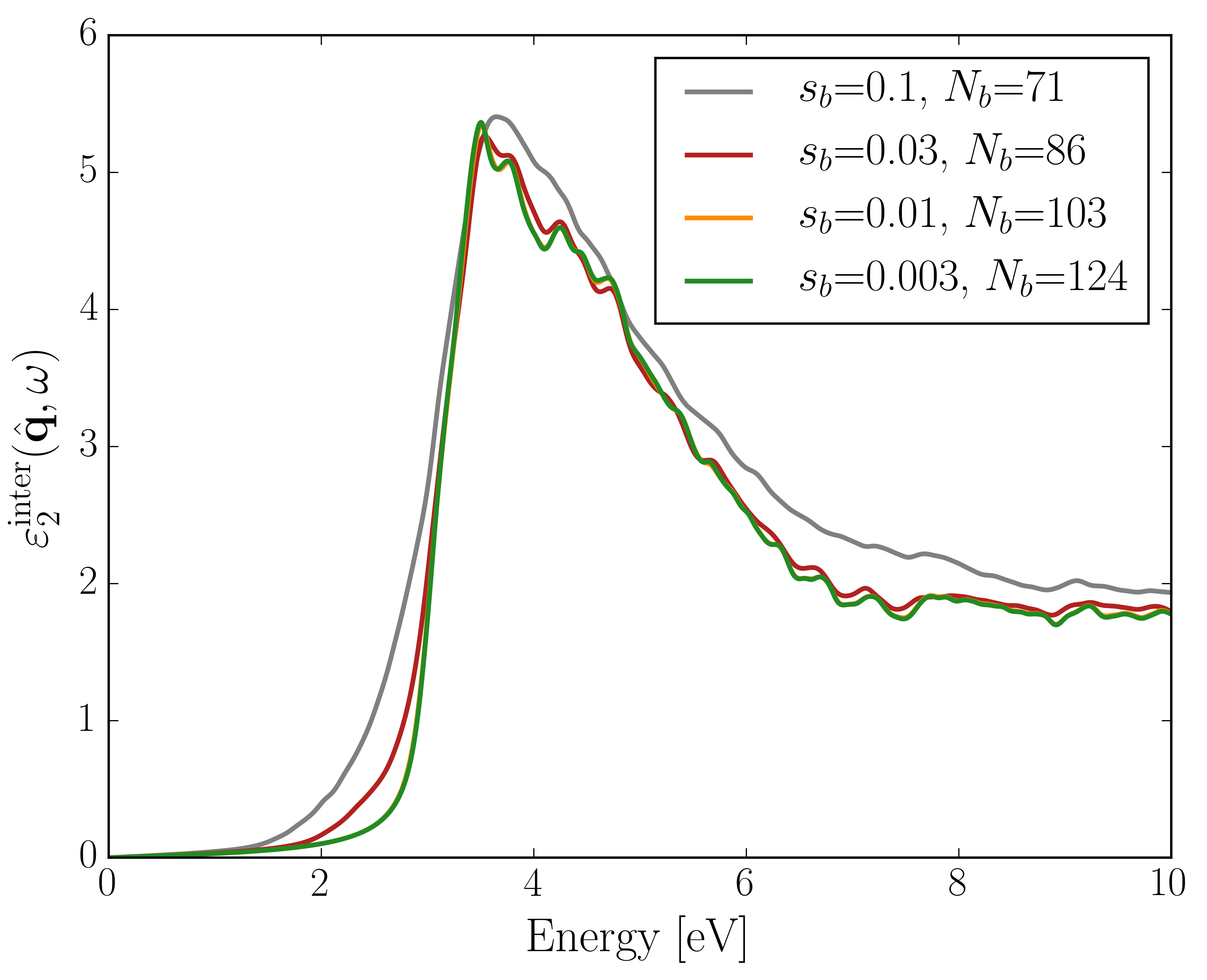}
\caption{Convergence of the IP complex dielectric function (here we show the imaginary part of the interband contribution $\Im{\left[\varepsilon_{\text{IP}}^{\text{inter}}(\hat{\mathbf{q}},\omega)\right]}$) for bulk Ag with respect to the size of the optimal basis, which is controlled by the $s_b$ threshold (\texttt{s\_bands} input variable for \texttt{simple.x}, in bohr$^3$). The corresponding dimension $N_b$ of the optimal basis is also reported.
\label{fig: IP_sb_convergence}}
\end{figure}

\begin{figure}[hbtp]
\centering
\includegraphics[width=\textwidth]{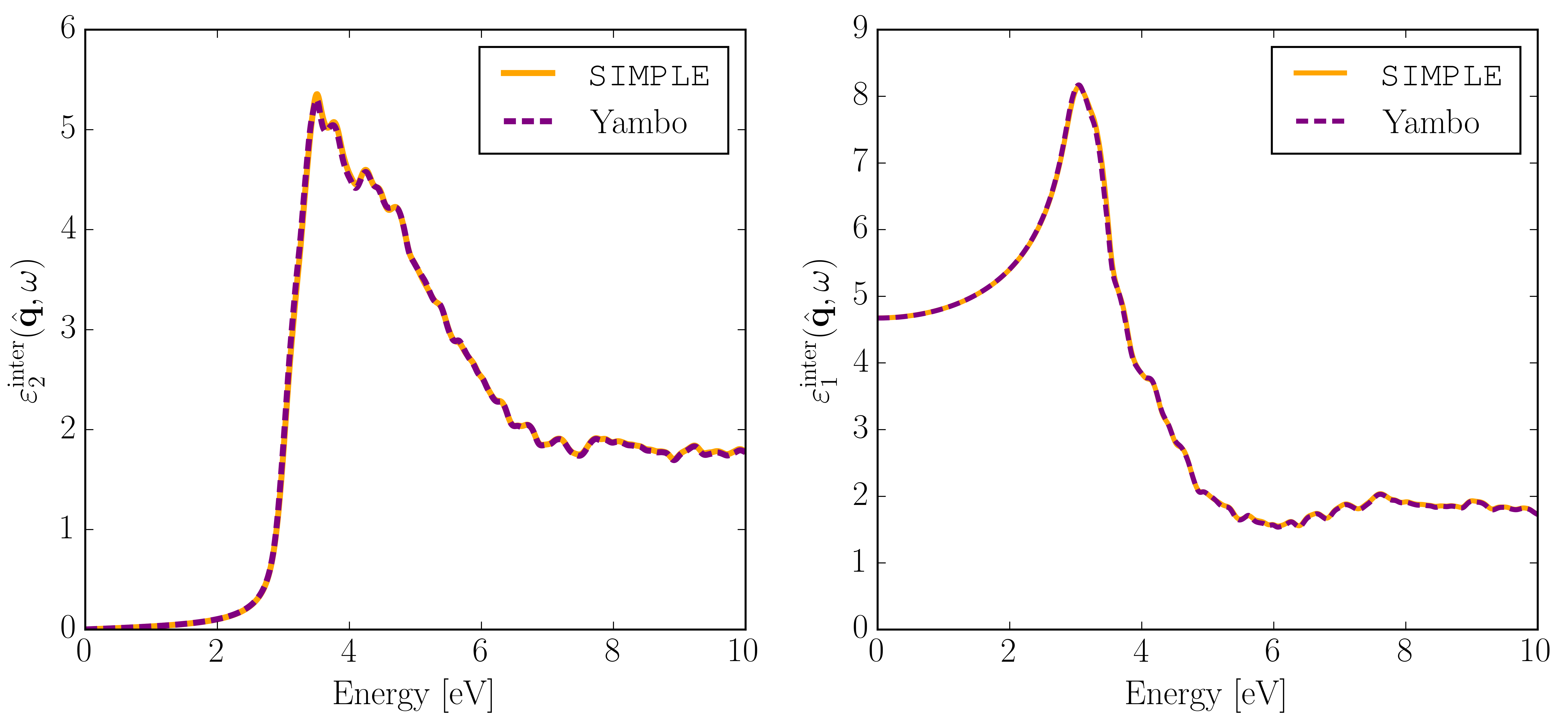}
\caption{Comparison of the interband contribution to the IP complex dielectric function of bulk Ag calculated with the \texttt{SIMPLE} code and with the Yambo code. We show both the real (right panel) and imaginary parts (left panel) of $\varepsilon_{\text{IP}}^{\text{inter}}(\hat{\mb{q}},\omega)$.  \label{fig: simple_ip-vs-yambo}}
\end{figure}

\noi
We consider as test system for the IP implementation bulk silver in the FCC primitive cell with lattice parameter $a=7.869$ bohr. The DFT calculations with the \texttt{pw.x} code of QE are performed with the PBE approximation~\cite{Perdew1996} for the exchange-correlation functional at a wavefunction cutoff of 55 Ry and using the scalar-relativistic norm-conserving pseudopotential from the SG15 library~\cite{Schlipf2015}. For the ground-state calculation (\textit{scf} calculation in \texttt{pw.x}) we use a $24 \times 24 \times 24$ Monkhorst-Pack~\cite{MonkhorstPack} k-grid while the periodic functions $\{ u_{n\mb{k}} \}$ needed for the construction of the OB are obtained on a $2 \times 2 \times 2$ uniform k-grid including the seven periodic images of $\Gamma$ and considering 11 conduction bands\footnote{Convergence of the optical spectra with respect to the \textit{nscf} k-grid has also been studied. In a primitive cell, a $2 \times 2 \times 2$ k-grid is more than enough to obtain converged results (see Supplementary Material) while in a supercell the $\Gamma$ point is already sufficient (in both cases including the seven periodic images of $\Gamma$).} (\textit{nscf} calculation in \texttt{pw.x}). The IP optical properties are calculated on a $44 \times 44 \times 44$ uniform k-grid with the inclusion of the non-local contribution to the velocity matrix elements and broadening parameters (i.e. \texttt{inter\_broadening}, \texttt{intra\_broadening} and \texttt{drude\_degauss} of Table~\ref{tab: input_simpleip}) of 0.1 eV. Because we are considering a FCC cubic crystal, the dielectric function is isotropic and is simply a scalar (i.e. independent from the direction $\hat{\mb{q}}$). The same holds for the test example of the BSE implementation, i.e. bulk silicon, which has the diamond structure.\\
First, we investigate the convergence of the IP dielectric function with respect to the threshold $s_{b}$ which controls the quality of the OB used for representing the periodic part of the Bloch wavefunctions. In Fig.~\ref{fig: IP_sb_convergence} we show how $\Im{\left[\varepsilon_{\text{IP}}^{\text{inter}}(\hat{\mb{q}},\omega)\right]}$ for Ag is fully converged at $s_b = 0.01$ bohr$^3$, or equivalently with $N_b=103$ basis functions (to be compared with the value $N_b=315$ corresponding to the complete basis, i.e. to $s_b=0.0$ bohr$^3$). Similar results are found for the convergence of $\Re{\left[\varepsilon_{\text{IP}}^{\text{inter}}(\hat{\mb{q}},\omega)\right]}$ and $\omega_{\text{D}}(\hat{\mb{q}})$, and are not shown here (see Supplementary Material). \\
Then, we verify the IP implementation of the \texttt{SIMPLE} code by comparing the results obtained for Ag with respect to equivalent simulations performed with the Yambo code~\cite{Marini2009} for the interband contribution to $\varepsilon_{\text{IP}}(\hat{\mb{q}},\omega)$, and with the BoltzWann module~\cite{Pizzi2014} of the Wannier90 code~\cite{Mostofi2008} for the intraband contribution to $\varepsilon_{\text{IP}}(\hat{\mb{q}},\omega)$. As both codes are interfaced with the QE distribution we can perform a straightforward comparison starting from the same DFT ground-state density used for the \texttt{SIMPLE} calculation using all the same relevant computational parameters (k-grid, number of bands, broadenings, etc.). To note that, in order to calculate the optical properties, Yambo does not use any interpolation method in k-space while BoltzWann, instead, uses the Wannier's interpolation method~\cite{Marzari2012}. For the \texttt{SIMPLE} calculation we use an OB constructed setting $s_b=0.01$ bohr$^3$ that, as shown above, gives well converged spectra. 
Fig.~\ref{fig: simple_ip-vs-yambo} shows that \texttt{SIMPLE} and Yambo give almost identical results for $\varepsilon_{\text{IP}}^{\text{inter}}(\hat{\mb{q}},\omega)$. For the intraband contribution we find that the Drude plasma frequency of Ag is $\omega_{\text{D}}(\hat{\mb{q}}) = 8.93$ eV with the \texttt{SIMPLE} code while it is $\omega_{\text{D}}(\hat{\mb{q}}) = 8.96$ eV with the BoltzWann code. 
Therefore we conclude that for both the interband and intraband contributions the agreement between SIMPLE and the other codes is excellent. \\
It is worth mentioning here that a powerful approach for the interpolation of band-structure properties is the interpolation mentioned above based on the use of maximally localized Wannier functions (MLWF) as basis~\cite{Marzari1997}. Wannier functions are physically appealing since they are localized in real space and can be used as an exact tight-binding basis. Moreover, Wannier interpolation is both an efficient and precise method for the evaluation of quantities that require a fine sampling of the BZ; in particular, in the Wannier representation the k-dependent KS Hamiltonian does not need to be explicitly constructed and the matrix elements of the velocity operator are analytic if written in terms of Wannier functions~\cite{Yates2007}. On the other hand, the method has the drawback that the construction of the MLWF's is not an easily automatizable procedure, especially for metals, for which an additional disentanglement step of the empty bands is required~\cite{Souza2001}. However, in recent years there have been promising advances towards the development of an automatic procedure for the ``Wannierization" of periodic systems~\cite{Damle2017}, including also the more complex case of metals~\cite{Damle2018}.
Once these new approaches will be interfaced with DFT engines, it would be interesting to systematically assess and compare performance and precision of Shirley's and Wannier's interpolations.

\subsection{BSE} \label{BSEresults}
\begin{figure}[hbtp]
\centering
\includegraphics[width=11cm]{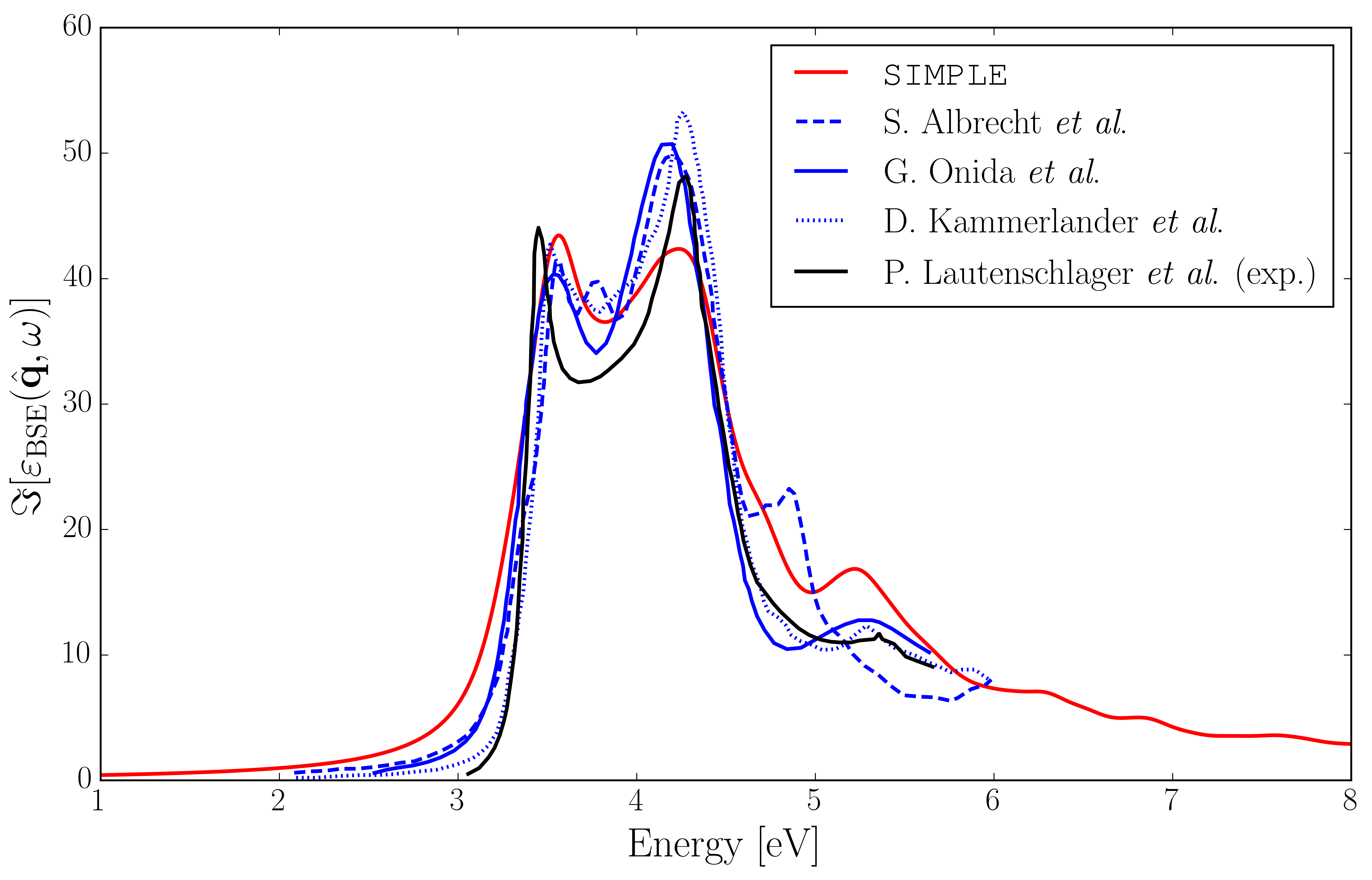}
\caption{Comparison of the imaginary part of the BSE complex dielectric function $\Im{\left[\varepsilon_{\text{BSE}}(\hat{\mb{q}},\omega)\right]}$ for bulk Si obtained with the \texttt{SIMPLE} code (solid red) with respect to previous BSE calculations: S. Albrecht \textit{et al}.~\cite{Albrecht} (dashed blue), G. Onida \textit{et al}.~\cite{Onida2002} (solid blue), D. Kammerlander \textit{et al}.~\cite{Kammerlander12} (dotted blue) and with respect to experiments (exp.): P. Lautenschlager \textit{et al}.~\cite{Lautenschlager87} (solid black).
\label{fig: BSE_Si_validation}}
\end{figure}

\noi
We test the BSE branch of the code considering bulk silicon as this was used, historically,  as a test case for the introduction of a variety of BSE implementations~\cite{Albrecht98,Onida2002,Rohlfing2000,Kammerlander12}. As the correlation part of the screened Coulomb interaction  is taken from a \texttt{GWL} calculation, performed at the sole $\Gamma$-point and then extrapolated as described in Section~\ref{section: implementation_bse}, we study a $8$ atoms cubic supercell. We work with the local-density  approximation~\cite{PZ81} (LDA) for the DFT exchange-correlation functional, with a norm-conserving pseudopotential (Si.pz-vbc.UPF) from the QE library and with a wavefunction cutoff of $20$ Ry for defining the plane-wave basis sets for the KS wavefunctions. These parameters yield a theoretical lattice constant of $10.26$ bohr. While for the starting DFT self-consistent run we use a regular $4\times 4 \times 4$ mesh k-points, for the convergence tests of the BSE implementation we use  a regular $6\times 6 \times 6$ mesh of k-points, shifted not to include the $\Gamma$-point. We include in the BSE simulations all the $16$ valence states, and $24$ conduction states.
We focus on the complex dielectric function which is obtained using a rigid shift of the valence band manifold, with a scissor parameter of $0.6$ eV for opening the electronic gap, as done in previous studies~\cite{Albrecht98,Rohlfing2000,Onida2002}, $1000$ iterative steps for the Haydock's recursive algorithm and a broadening parameter $\eta$ of $0.2$ eV. We note that using only  $\sim 50$   steps is enough to ensure convergence.
A polarizability basis~\cite{Umari09,Umari2010} of $100$ vectors is enough to converge the screened Coulomb interaction (see Supplementary Material).
We show in Fig.~\ref{fig: BSE_sb-sp_convergence} (left panel) the calculated $\Im{\left[\varepsilon_{\text{BSE}}(\hat{\mb{q}},\omega)\right]}$ as a function of the $s_b$ threshold that controls the quality of the OB: a value of $0.1$ bohr$^3$ is sufficient for converging the spectrum, so that the OB comprises only 91 vectors. This number, 8640, should be contrasted with the theoretical dimension of a complete basis (i.e. $s_b=0.0$ bohr$^3$), and with the dimension of the plane-wave basis sets for representing the KS wavefunctions, that has 2205 elements.
We show in Fig.~\ref{fig: BSE_sb-sp_convergence} (right panel) how $\Im{\left[\varepsilon_{\text{BSE}}(\hat{\mb{q}},\omega)\right]}$ behaves if we lower the $s_p$ threshold controlling the quality of the OPB. For this test the OB threshold is kept fixed to $s_b=0.1$ bohr$^3$ giving  a basis dimension $N_b=91$. We see how the choice $s_p=1.0$ bohr$^3$ provides a spectrum almost indistinguishable from the one obtained with $s_p=0.5$ bohr$^3$, while the dimension of the OPB raises from $N_p=160$ to $N_p=443$.
From our tests of the sensitivity of the dielectric function on the $s_b$ and $s_p$ thresholds we opt for the choice $s_b=0.1$ bohr$^3$ and $s_p=1.0$ bohr$^3$ for verifying our BSE implementation. As the spectra are quite sensitive to k-points sampling~\cite{albrecht_reply, Kammerlander12}, we take a denser k-point mesh comprising $448$ points. We compare in Fig.~\ref{fig: BSE_Si_validation} the resulting $\Im{\left[\varepsilon_{\text{BSE}}(\hat{\mb{q}},\omega)\right]}$ spectrum with those taken from Refs.~\cite{Albrecht98,Rohlfing2000,Onida2002,Kammerlander12}, registering good agreement despite the different approximations the various implementations rely on. In particular, the strategy used for extrapolating the $W_c$ operator is verified.

\section{Performance}

\subsection{Efficiency of the optimal basis}
\noi
As the core of the present approach is the reduction of the basis sets for representing wavefunctions and their products, we are compelled to assess how such reduction improves the computational performance. 
\subsubsection{IP}
\begin{figure}[hbtp]
\centering
\includegraphics[width=10cm]{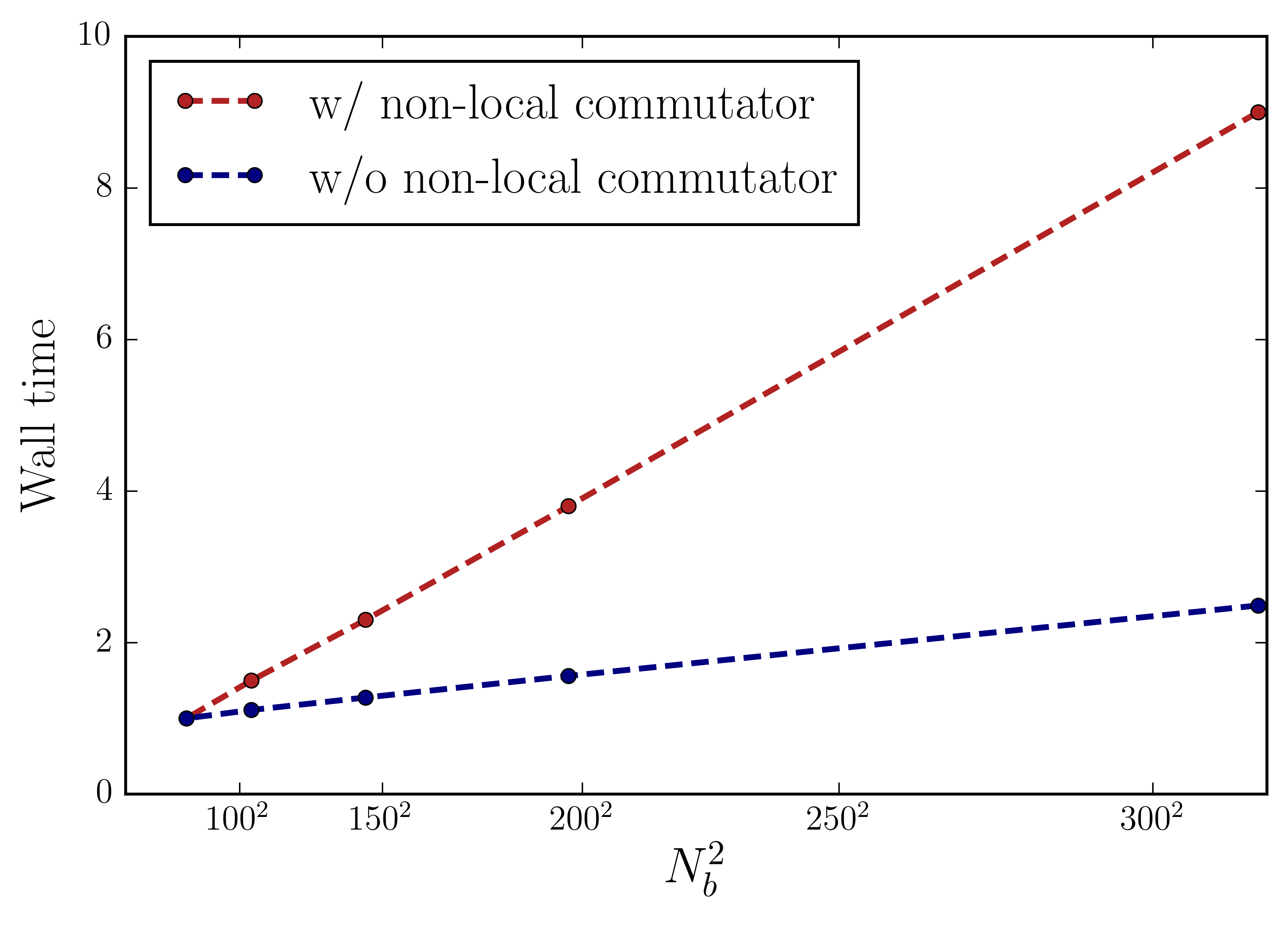}
\caption{Relative computational time for a IP calculation performed with the \texttt{SIMPLE} code on bulk Ag, both including and not including the non-local commutator in the evaluation of the velocity matrix elements,  with respect to the square of the number of optimal basis functions $N_b$. The reference wall time is set as the one resulting from the calculation with the smallest number of vectors. 
\label{fig: speedup_IP_basis}}
\end{figure}

\noi 
For IP calculations we have to deal with matrices of dimensions $N_b \times N_b$ (i.e. the k-dependent Hamiltonian $H^{\text{KS}}_{ij}(\mb{k})$ of Eq.~\ref{eq: H_ij_k}) and the computational cost is dictated by the construction (or ``filling") of this matrix through the computation of the k-dependent contributions coming from the non-local part of the pseudopotentials, rather than by its diagonalization.  For this reason the wall time of a IP calculation is expected to scale quadratically with respect to $N_b$. \\
We study the computational cost of the IP approach performed with the
\texttt{SIMPLE} code as the number of OB functions $N_b$ is varied, as shown in Fig.~\ref{fig: speedup_IP_basis} for bulk silver. We set as reference time the computer time (or wall time) of the calculation with the smallest number of OB functions. As expected, timing scales with the square of $N_b$. In Fig.~\ref{fig: speedup_IP_basis} we also show how the inclusion of the non-local commutator in the computation of the velocity matrix elements significantly slows down the calculations.
\subsubsection{BSE}
\begin{figure}[hbtp]
\centering
\includegraphics[width=\textwidth]{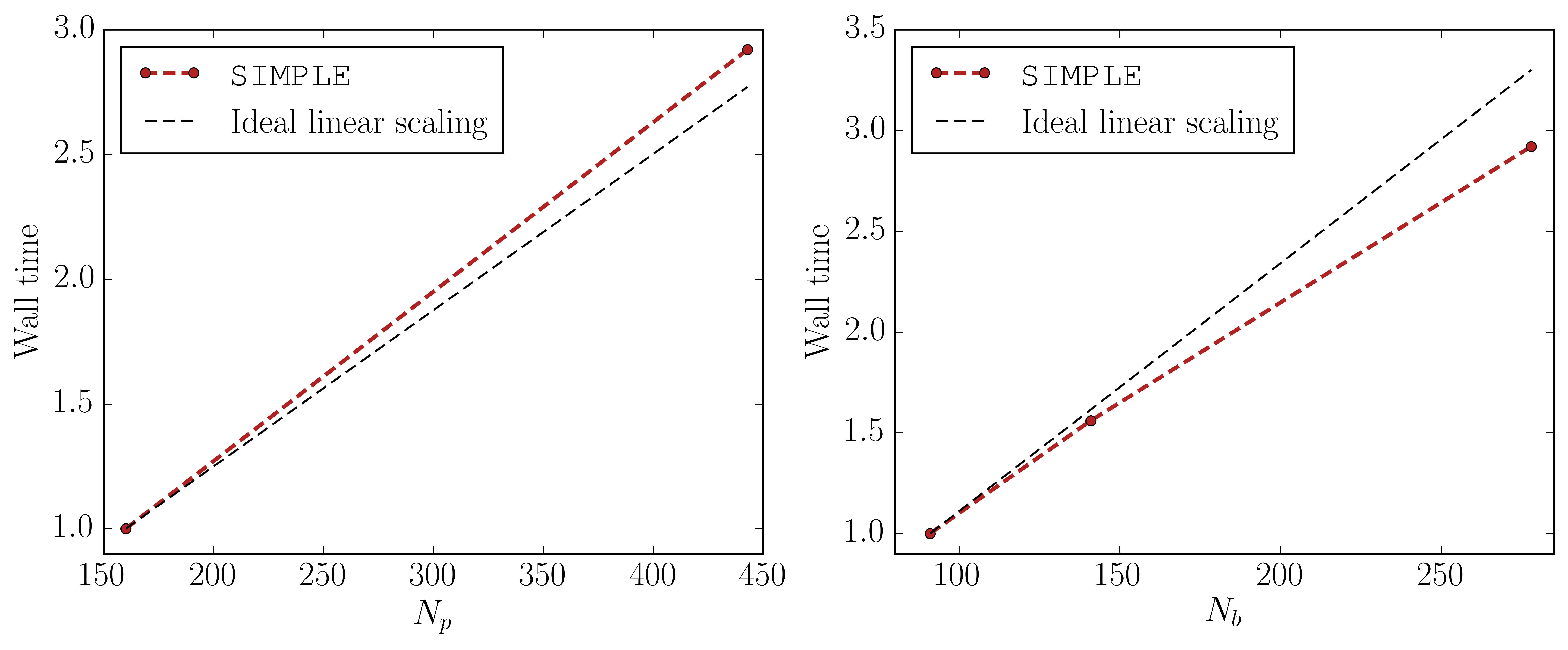}
\caption{Relative computational time for a BSE calculation performed with the \texttt{SIMPLE} code on bulk Si with respect to the number of optimal product basis functions $N_p$ (left panel) and to the number of optimal basis functions $N_b$ (right panel). The reference wall time is set as the one resulting from the calculation with the smallest number of vectors.  \label{fig: BSE_scaling_basis_nb-np}}
\end{figure}

\noi
 For BSE calculations  we work with matrices of dimensions $N_b \times N_v$, $N_b \times N_c$ and $N_p \times N_\pi$ where $N_{\pi}$ is the dimension of the basis used to represent the polarizability operator (see Eq.~\ref{eq: W_c_GWW}). As $N_{\pi}$ does not depend on OB or OPB, we expect the wall time of a BSE calculation to scale linearly with respect to $N_p$.  Besides, we also know that the number of vectors of the OPB $N_p$ scales linearly with $N_b$, instead of its square, because of the  threshold  $s_p$ used in the construction of the basis (see Section~\ref{section: implementation_bse}). As a consequence, the wall time of a BSE calculation is also expected to scale linearly with respect to $N_b$. \\
We show in Fig.~\ref{fig: BSE_scaling_basis_nb-np} (left panel) how the total computational cost depends almost linearly on the dimension $N_p$ of the OPB, as expected (with the $s_b$ threshold kept fixed).  Indeed, we register a scaling better than linear likely due to a more efficient use of the cache memory. In Fig.~\ref{fig: BSE_scaling_basis_nb-np} (right panel) we show that linear scaling is also found for the variation of the $N_b$ size of the OB (with the $s_p$ threshold kept fixed). In this case the slightly worse than ideal behavior should be traced to the requirement of storing and working with larger matrices, implying significant transfers from  memory to CPU and back.

\subsection{Core scaling}
\noi 
In order to assess the performances of the \texttt{SIMPLE} code for its use in parallel architectures, we investigate the speedup of the code as a function of the number of single processing units, or cores, both for the IP and BSE implementation.
\subsubsection{IP}
\begin{figure}[hbtp]
\centering
\includegraphics[width=10.5cm]{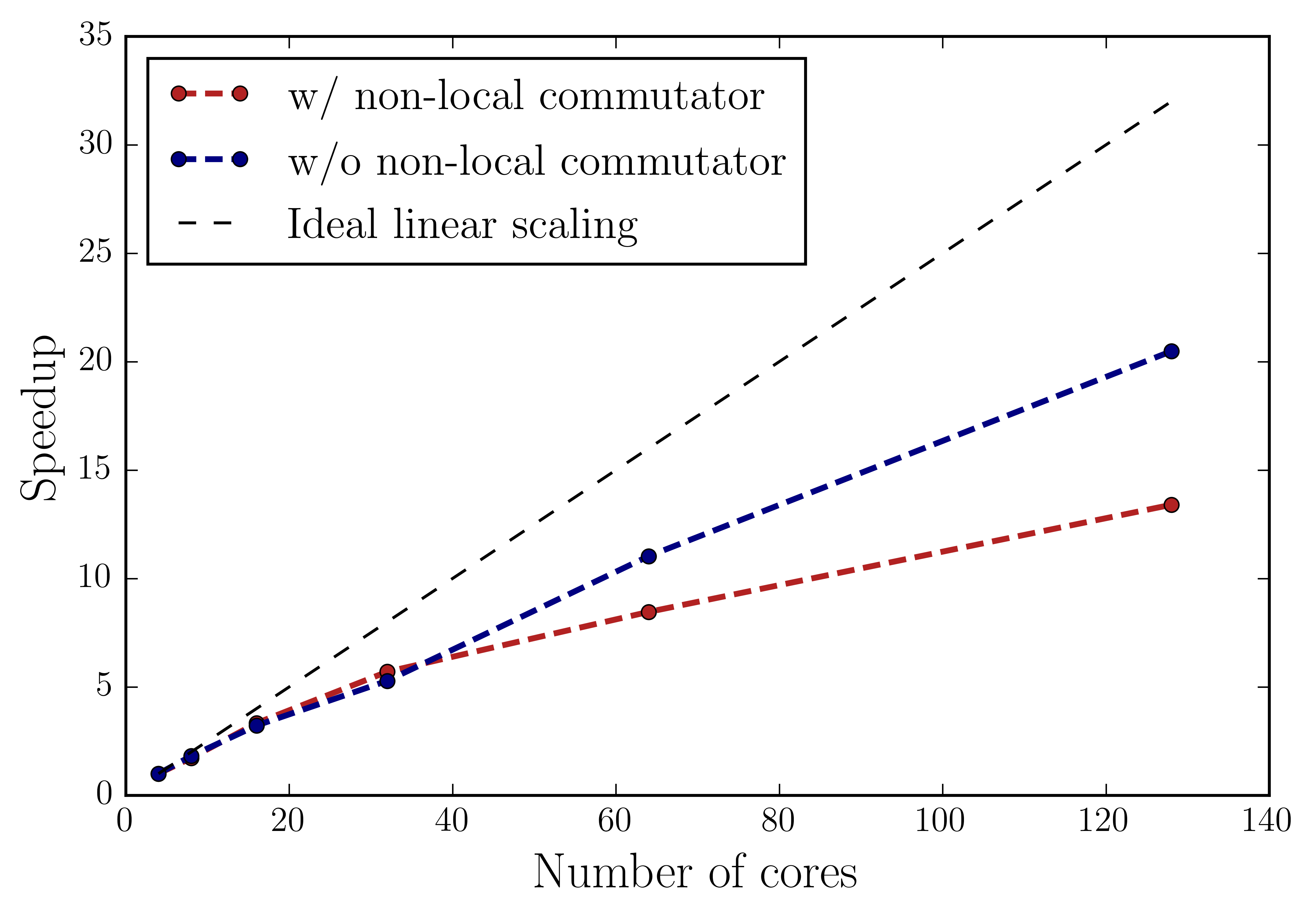}
\caption{Speedup of an IP calculation performed with the \texttt{SIMPLE} code (\texttt{simple.x} and \texttt{simple\_ip.x}) as a function of the number of cores for bulk Ag in a $4 \times 4 \times 4$ supercell with 64 atoms including and not including the contribution to the velocity matrix elements coming from the non-local commutator. The test is performed on Intel\textsuperscript{\textregistered} Skylake\textsuperscript{\textregistered} processors.
\label{fig: speedup_IP_cpu}}
\end{figure}

\noi 
For the IP part we consider Ag in a $4 \times 4 \times 4$ FCC supercell with 64 atoms and 1216 electrons. The total number of OB functions considered in the simulations is 2687, corresponding to $s_b=0.01$ bohr$^3$. 
In general, for IP calculations, a significant part of the total computational cost is due to the \texttt{simple.x} run. Therefore the speedup of the IP section is calculated by summing the wall times of the \texttt{simple.x} and \texttt{simple\_ip.x} runs performed with the same number of cores.
In Fig.~\ref{fig: speedup_IP_cpu} we show the results for the case in which the number of cores corresponds to the number of MPI tasks and we compare simulations performed including and not including the computation of the non-local commutator entering in the estimation of the velocity matrix elements. In order to obtain similar wall times for a given number of cores, we use an interpolation k-grid of $8 \times 8 \times 8$ and $16 \times 16 \times 16$ when including and not including the non-local commutator, respectively. 
Although the scaling of \texttt{simple\_ip.x} alone is close to linear, the total scaling of the combined simulations deviates from the ideal linear scaling, mainly because of the computations of the k-dependent matrix elements due to the non-local part of the pseudopotentials (see Eq.~\ref{eq: Vnloc_ij_k} and Eq.~\ref{eq: velocity_nn'_k}) performed by \texttt{simple.x}.  Indeed, if the calculation of the non-local commutator of Eq.~\ref{eq: velocity_nn'_k} is avoided, the scaling is significantly improved, as shown in Fig.~\ref{fig: speedup_IP_cpu}. 
\subsubsection{BSE}
\begin{figure}[hbtp]
\centering
\includegraphics[width=\textwidth]{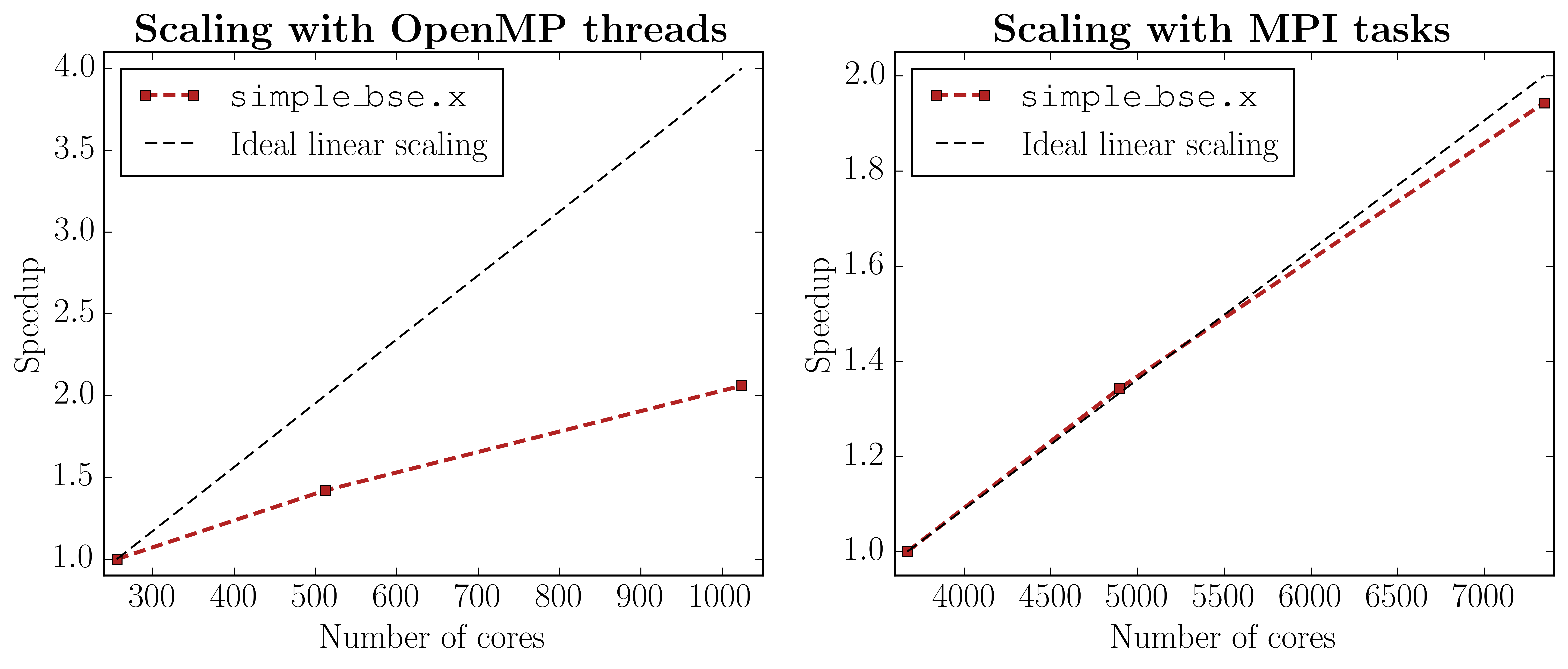}
\caption{Speedup of a BSE calculation performed with the \texttt{SIMPLE} code (\texttt{simple\_bse.x}) as a function of the number of computing cores for bulk CH$_3$NH$_3$PbI$_3$ (48 atoms) with a $4\times 4\times 4$ mesh of k-points. The cost of the foregoing \texttt{simple.x} run is negligible. (Left panel) The number of MPI tasks is kept fixed to 64 while the number of OpenMP threads for each task varies from 4 to 16. (Right panel) The number of OpenMP threads for each MPI task is kept fixed to 68 while the number of the latter varies from 54 to 108. The test is performed on Intel\textsuperscript{\textregistered} Xeon Phi7250\textsuperscript{\textregistered} (KnightLandings\textsuperscript{\textregistered}) processors.  \label{fig: BSE_scaling_cpu_openmp-mpi}}
\end{figure}

\noi
For the BSE part, in order to illustrate the capabilities of the code, we choose the mixed organic-inorganic perovskite CH$_3$NH$_3$PbI$_3$ which is widely studied for organic/inorganic perovskite solar cells~\cite{Umari14}. It involves a primitive cell, comprising 48 inequivalent atoms, fully relativistic calculations and 200 electron. We test the evaluation of the dielectric function through the Haydock's recursive algorithm. The code is optimized for a mixed OpenMP-MPI parallelization as it strongly relies on BLAS routines and in particular on the DGEMM matrix-matrix multiplication.
First, we check the scaling with respect to the number of OpenMP threads keeping fixed the number of MPI tasks. The corresponding results are displayed in Fig.~\ref{fig: BSE_scaling_cpu_openmp-mpi} (left panel), where we do not consider the cost of the preceding \texttt{simple.x} run as this is smaller: $3369$ s of wall time on only 128 cores, to be contrasted with $8782$ s for the  \texttt{simple\_bse.x} run with 256 cores. We note that changing from 4 to 16 the number of OpenMP threads per MPI task determines a reduction of the wall time by a factor $2.1$. Although this figure is lower than the ideal factor $4$, we register an almost linear scaling with the number of OpenMP threads. Indeed, an only slightly larger factor ($2.3$) would be expected extrapolating the scaling from 256 to 512 cores. Such linear scaling supports the parallelization of the single MPI tasks on a relatively large number of OpenMP threads. 
The scaling of the required wall time with respect to the number of MPI tasks keeping fixed the distribution of the single tasks on OpenMP threads is displayed in  Fig.~\ref{fig: BSE_scaling_cpu_openmp-mpi} (right panel). Also in this test we focus only on the \texttt{simple\_bse.x} run  as the \texttt{simple.x} one is much less demanding ($7840$ s of wall time on only $128$ cores versus $12251$ s for the \texttt{simple\_bse.x} calculation on $3672$ cores). The number of OpenMP threads per MPI task is kept fixed to $68$ while the number of MPI tasks varies from $54$ to $108$.
We register linear scaling and, in this case, we note also an almost ideal behavior. 
The favorable scaling both with respect to the number of OpenMP threads and of MPI tasks will allow the use of our BSE code on the latest high-performance computing infrastructures running on up to few tens of thousands of cores.

\section{Conclusions}
\noi 
We have developed a new code, named \texttt{SIMPLE}, to calculate optical properties using the Shirley's optimal basis method proposed in Ref.~\cite{Shirley1996}, both within the framework of the independent-particle approximation (for metals) and of the Bethe-Salpeter equation (for insulators). \\
In the independent-particle implementation, we highlight that the code computes both interband and intraband contributions to the complex dielectric function and that the matrix elements of the velocity operator are correctly calculated by including, if desired, the non-local contribution of the norm-conserving pseudopotentials. 
In the Bethe-Salpeter part, a generalization of the optimal basis method has been developed and implemented to deal in an efficient way also with products of periodic Kohn-Sham states. 
The screened Coulomb interaction is obtained either through a model for the microscopic dielectric function or starting from the results of a previous \texttt{GWL} calculation.
To avoid the diagonalization of the excitonic Hamiltonian matrix, the lowest eigenvectors and corresponding energies can be obtained exploiting an iterative conjugate-gradient algorithm while the complex dielectric function is calculated with the Haydock's recursive method.\\
The \texttt{SIMPLE} code has been included within the \texttt{GWL} package of the Quantum ESPRESSO distribution. It has been developed and optimized, as demonstrated by scalability tests, in order to run efficiently in laptops as well as in massively parallel architectures by deploying both fast linear algebra routines (BLAS, LAPACK) and parallelization strategies (MPI and OpenMP). 
We have verified the results of our code obtained within the framework of  the independent-particle approximation and of the Bethe-Salpeter equation on bulk silver and bulk silicon, respectively.
In the case of silver we have compared \texttt{SIMPLE} with the publicly available codes Yambo and BoltzWann and have found very good agreement. In the case of silicon, for which the Bethe-Salpeter equation is solved, \texttt{SIMPLE} has been compared with the literature, finding also here a good agreement for the results, considering the use of different base codes and pseudopotentials.\\ 
The systematic improvement of the optimal basis set and, as a consequence, of the precision of the results by simply increasing the number of basis functions included, provides the notable advantage that our code can be used to perform efficient computations in a straightforward way for the users:  precision is controlled simply by a few input parameters, i.e. $s_b$ and $s_p$.  
Moreover, the simplicity of the method and the fact that all the calculations needed to obtain the optical properties from a given initial crystal structure can be run within a single open-source software, i.e. the Quantum ESPRESSO distribution, makes the automation of the sequence of computational steps required and thus the systematic evaluation of the complex dielectric function for a large number of materials much easier.
Because of all these considerations, we hope that the \texttt{SIMPLE} code will have an impact in the field of computational materials science by helping researchers to perform efficient, systematic, and reliable first-principles studies of the optical properties of materials.\\

\section*{Acknowledgments}
\noi
G. P. acknowledges funding from Varinor SA (CH 2800 Del\'emont, Switzerland).
P. U. acknowledges funding from the EU-H2020 research and innovation programme under Grant Agreement No. 654360 NFFA-Europe.
This research was partially supported by the NCCR MARVEL, funded by the Swiss National Science Foundation.
We acknowledge PRACE (project ID 2016163963) for awarding us access to Marconi at CINECA, Italy. 
G. P. warmly thanks Gian-Marco Rignanese for originally suggesting the use of the optimal basis method.

\section*{Declarations of interest: none}

\section*{References}





\bibliographystyle{elsarticle-num}

\begin{thebibliography}{10}
\expandafter\ifx\csname url\endcsname\relax
  \def\url#1{\texttt{#1}}\fi
\expandafter\ifx\csname urlprefix\endcsname\relax\def\urlprefix{URL }\fi
\expandafter\ifx\csname href\endcsname\relax
  \def\href#1#2{#2} \def\path#1{#1}\fi

\bibitem{Martin2004}
R.~M. Martin,
  \href{https://www.cambridge.org/core/books/electronic-structure/DDFE838DED61D7A402FDF20D735BC63A}{{Electronic
  Structure: Basic Theory and Practical Methods}}, Cambridge University Press,
  2004.

\bibitem{Strinati1988}
G.~Strinati,
  \href{https://link.springer.com/article/10.1007%2FBF02725962}{{Application of
  the Green's Functions Method to the Study of the Optical Properties of
  Semiconductors}}, Riv. Nuovo Cimento 11~(12) (1988) 1--86.

\bibitem{Onida2002}
G.~Onida, L.~Reining, A.~Rubio,
  \href{https://journals.aps.org/rmp/abstract/10.1103/RevModPhys.74.601}{{Electronic
  excitations: density-functional versus many-body Green’s-function
  approaches}}, Rev. Mod. Phys. 74 (2002) 601--659.

\bibitem{Martin2016}
R.~Martin, L.~Reining, D.~Ceperley,
  \href{https://books.google.ch/books?id=ch1CDAAAQBAJ}{{Interacting Electrons:
  Theory and Computational Approaches}}, Cambridge University Press, 2016.

\bibitem{Hybertsen1986}
M.~S. Hybertsen, S.~G. Louie,
  \href{https://link.aps.org/doi/10.1103/PhysRevB.34.5390}{Electron correlation
  in semiconductors and insulators: Band gaps and quasiparticle energies},
  Phys. Rev. B 34 (1986) 5390--5413.

\bibitem{Dabo2010}
I.~Dabo, A.~Ferretti, N.~Poilvert, Y.~Li, N.~Marzari, M.~Cococcioni,
  \href{https://link.aps.org/doi/10.1103/PhysRevB.82.115121}{Koopmans'
  condition for density-functional theory}, Phys. Rev. B 82 (2010) 115121.

\bibitem{Borghi2014}
G.~Borghi, A.~Ferretti, N.~L. Nguyen, I.~Dabo, N.~Marzari,
  \href{https://link.aps.org/doi/10.1103/PhysRevB.90.075135}{Koopmans-compliant
  functionals and their performance against reference molecular data}, Phys.
  Rev. B 90 (2014) 075135.

\bibitem{Nguyen2018}
N.~L. Nguyen, N.~Colonna, A.~Ferretti, N.~Marzari,
  \href{https://link.aps.org/doi/10.1103/PhysRevX.8.021051}{Koopmans-compliant
  spectral functionals for extended systems}, Phys. Rev. X 8 (2018) 021051.

\bibitem{Colonna2018}
N.~Colonna, N.~L. Nguyen, A.~Ferretti, N.~Marzari,
  \href{https://doi.org/10.1021/acs.jctc.7b01116}{Screening in
  orbital-density-dependent functionals}, J. Chem. Theory Comput. 14~(5) (2018)
  2549--2557.

\bibitem{Marzari2016}
N.~Marzari, \href{https://doi.org/10.1038/nmat4613}{{Materials modelling: The
  frontiers and the challenges}}, Nat. Mater. 15 (2016) 381.

\bibitem{Marini2009}
A.~Marini, C.~Hogan, M.~Gr{\"{u}}ning, D.~Varsano,
  \href{http://www.sciencedirect.com/science/article/pii/S0010465509000472}{yambo:
  An \emph{ab initio} tool for excited state calculations}, Comput. Phys.
  Commun. 180~(8) (2009) 1392--1403.

\bibitem{Deslippe2012}
J.~Deslippe, G.~Samsonidze, D.~A. Strubbe, M.~Jain, M.~L. Cohen, S.~G. Louie,
  \href{http://www.sciencedirect.com/science/article/pii/S0010465511003912}{{BerkeleyGW:
  A massively parallel computer package for the calculation of the
  quasiparticle and optical properties of materials and nanostructures}},
  Comput. Phys. Commun. 183~(6) (2012) 1269--1289.

\bibitem{Leng2016}
X.~Leng, F.~Jin, M.~Wei, Y.~Ma,
  \href{https://onlinelibrary.wiley.com/doi/full/10.1002/wcms.1265}{{GW method
  and Bethe-Salpeter equation for calculating electronic excitations}}, WIREs
  Comput. Mol. Sci. 6 (2016) 532--550.

\bibitem{Giannozzi2009}
P.~Giannozzi, S.~Baroni, N.~Bonini, M.~Calandra, R.~Car, C.~Cavazzoni,
  D.~Ceresoli, G.~L. Chiarotti, M.~Cococcioni, I.~Dabo, A.~{Dal Corso},
  S.~de~Gironcoli, S.~Fabris, G.~Fratesi, R.~Gebauer, U.~Gerstmann,
  C.~Gougoussis, A.~Kokalj, M.~Lazzeri, L.~Martin-Samos, N.~Marzari, F.~Mauri,
  R.~Mazzarello, S.~Paolini, A.~Pasquarello, L.~Paulatto, C.~Sbraccia,
  S.~Scandolo, G.~Sclauzero, A.~P. Seitsonen, A.~Smogunov, P.~Umari, R.~M.
  Wentzcovitch, \href{dx.doi.org/10.1088/0953-8984/21/39/395502}{{QUANTUM
  ESPRESSO: a modular and open-source software project for quantum simulations
  of materials.}}, J. Phys.: Condens. Matter 21~(39) (2009) 395502.

\bibitem{Gonze2009}
X.~Gonze, B.~Amadon, P.-M. Anglade, J.-M. Beuken, F.~Bottin, P.~Boulanger,
  F.~Bruneval, D.~Caliste, R.~Caracas, M.~Côté, T.~Deutsch, L.~Genovese,
  P.~Ghosez, M.~Giantomassi, S.~Goedecker, D.~Hamann, P.~Hermet, F.~Jollet,
  G.~Jomard, S.~Leroux, M.~Mancini, S.~Mazevet, M.~Oliveira, G.~Onida,
  Y.~Pouillon, T.~Rangel, G.-M. Rignanese, D.~Sangalli, R.~Shaltaf, M.~Torrent,
  M.~Verstraete, G.~Zerah, J.~Zwanziger,
  \href{http://www.sciencedirect.com/science/article/pii/S0010465509002276}{{ABINIT:
  First-principles approach to material and nanosystem properties}}, Comput.
  Phys. Commun. 180~(12) (2009) 2582--2615.

\bibitem{Soler2002}
J.~M. Soler, E.~Artacho, J.~D. Gale, A.~Garc\'ia, J.~Junquera, P.~Ordej\'on,
  D.~S\'anchez-Portal,
  \href{http://stacks.iop.org/0953-8984/14/i=11/a=302}{{The SIESTA method for
  \textit{ab initio} order-N materials simulation}}, J. Phys.: Condens. Matter
  14~(11) (2002) 2745.

\bibitem{Kresse1996}
G.~Kresse, J.~Furthmüller,
  \href{http://www.sciencedirect.com/science/article/pii/0927025696000080}{Efficiency
  of ab-initio total energy calculations for metals and semiconductors using a
  plane-wave basis set}, Comput. Mater. Sci. 6~(1) (1996) 15--50.

\bibitem{Hutter2014}
J.~Hutter, M.~Iannuzzi, F.~Schiffmann, J.~VandeVondele,
  \href{https://onlinelibrary.wiley.com/doi/abs/10.1002/wcms.1159}{cp2k:
  atomistic simulations of condensed matter systems}, Wiley Interdiscip. Rev.
  Comput. Mol. Sci. 4~(1) (2014) 15--25.

\bibitem{Gonze2016}
X.~Gonze, F.~Jollet, F.~A. Araujo, D.~Adams, B.~Amadon, T.~Applencourt,
  C.~Audouze, J.-M. Beuken, J.~Bieder, A.~Bokhanchuk, E.~Bousquet, F.~Bruneval,
  D.~Caliste, M.~Côté, F.~Dahm, F.~D. Pieve, M.~Delaveau, M.~D. Gennaro,
  B.~Dorado, C.~Espejo, G.~Geneste, L.~Genovese, A.~Gerossier, M.~Giantomassi,
  Y.~Gillet, D.~Hamann, L.~He, G.~Jomard, J.~L. Janssen, S.~L. Roux, A.~Levitt,
  A.~Lherbier, F.~Liu, I.~Lukačević, A.~Martin, C.~Martins, M.~Oliveira,
  S.~Poncé, Y.~Pouillon, T.~Rangel, G.-M. Rignanese, A.~Romero, B.~Rousseau,
  O.~Rubel, A.~Shukri, M.~Stankovski, M.~Torrent, M.~V. Setten, B.~V. Troeye,
  M.~Verstraete, D.~Waroquiers, J.~Wiktor, B.~Xu, A.~Zhou, J.~Zwanziger,
  \href{http://www.sciencedirect.com/science/article/pii/S0010465516300923}{{Recent
  developments in the ABINIT software package}}, Comput. Phys. Commun. 205
  (2016) 106--131.

\bibitem{Giannozzi2017}
P.~Giannozzi, O.~Andreussi, T.~Brumme, O.~Bunau, M.~B. Nardelli, M.~Calandra,
  R.~Car, C.~Cavazzoni, D.~Ceresoli, M.~Cococcioni, N.~Colonna, I.~Carnimeo,
  A.~D. Corso, S.~de~Gironcoli, P.~Delugas, R.~A.~D. Jr, A.~Ferretti,
  A.~Floris, G.~Fratesi, G.~Fugallo, R.~Gebauer, U.~Gerstmann, F.~Giustino,
  T.~Gorni, J.~Jia, M.~Kawamura, H.-Y. Ko, A.~Kokalj, E.~Kucukbenli,
  M.~Lazzeri, M.~Marsili, N.~Marzari, F.~Mauri, N.~L. Nguyen, H.-V. Nguyen,
  A.~O. de-la Roza, L.~Paulatto, S.~Poncé, D.~Rocca, R.~Sabatini, B.~Santra,
  M.~Schlipf, A.~P. Seitsonen, A.~Smogunov, I.~Timrov, T.~Thonhauser, P.~Umari,
  N.~Vast, X.~Wu, S.~Baroni,
  \href{http://iopscience.iop.org/article/10.1088/1361-648X/aa8f79}{{Advanced
  capabilities for materials modelling with Quantum ESPRESSO}}, J. Phys.:
  Condens. Matter 29~(46) (2017) 465901.

\bibitem{Shirley1996}
E.~L. Shirley,
  \href{https://link.aps.org/doi/10.1103/PhysRevB.54.16464}{{Optimal basis sets
  for detailed Brillouin-zone integrations}}, Phys. Rev. B 54 (1996)
  16464--16469.

\bibitem{Yao2004}
Y.~Yao, L.~Kleinman, A.~H. MacDonald, J.~Sinova, T.~Jungwirth, D.-s. Wang,
  E.~Wang, Q.~Niu,
  \href{https://link.aps.org/doi/10.1103/PhysRevLett.92.037204}{{First
  Principles Calculation of Anomalous Hall Conductivity in Ferromagnetic bcc
  Fe}}, Phys. Rev. Lett. 92 (2004) 037204.

\bibitem{Sangalli2012}
D.~Sangalli, A.~Marini, A.~Debernardi,
  \href{https://link.aps.org/doi/10.1103/PhysRevB.86.125139}{{Pseudopotential-based
  first-principles approach to the magneto-optical Kerr effect: From metals to
  the inclusion of local fields and excitonic effects}}, Phys. Rev. B 86 (2012)
  125139.

\bibitem{Sangalli2017}
D.~Sangalli, J.~A. Berger, C.~Attaccalite, M.~Gr\"uning, P.~Romaniello,
  \href{https://link.aps.org/doi/10.1103/PhysRevB.95.155203}{{Optical
  properties of periodic systems within the current-current response framework:
  Pitfalls and remedies}}, Phys. Rev. B 95 (2017) 155203.

\bibitem{Gatti2013}
M.~Gatti, F.~Sottile,
  \href{https://link.aps.org/doi/10.1103/PhysRevB.88.155113}{Exciton dispersion
  from first principles}, Phys. Rev. B 88 (2013) 155113.

\bibitem{Wooten1972}
F.~Wooten, \href{https://books.google.ch/books?id=Vzl0oAEACAAJ}{Optical
  properties of solids}, Academic Press, 1972.

\bibitem{Marini2001}
A.~Marini, G.~Onida, R.~{Del Sole},
  \href{https://journals.aps.org/prb/abstract/10.1103/PhysRevB.64.195125}{{Plane-wave
  DFT-LDA calculation of the electronic structure and absorption spectrum of
  copper}}, Phys. Rev. B 64~(19) (2001) 195125.

\bibitem{Harl_phd}
J.~Harl, \href{http://othes.univie.ac.at/2622/}{{The linear response function
  in density functional theory: Optical spectra and improved description of the
  electron correlation}}, Ph.D. thesis, {Universit\"at Wien} (2008).

\bibitem{Mustafa2016}
J.~I. Mustafa, M.~Bernardi, J.~B. Neaton, S.~G. Louie,
  \href{https://link.aps.org/doi/10.1103/PhysRevB.94.155105}{Ab initio
  electronic relaxation times and transport in noble metals}, Phys. Rev. B 94
  (2016) 155105.

\bibitem{Albrecht}
S.~Albrecht, L.~Reining, R.~Del~Sole, G.~Onida,
  \href{https://journals.aps.org/prl/abstract/10.1103/PhysRevLett.80.4510}{{Ab
  Initio Calculation of Excitonic Effects in the Optical Spectra of
  Semiconductors}}, Phys. Rev. Lett. 80 (1998) 4510--4513.

\bibitem{Rohlfing2000}
M.~Rohlfing, S.~G. Louie,
  \href{https://journals.aps.org/prb/abstract/10.1103/PhysRevB.62.4927}{Electron-hole
  excitations and optical spectra from first principles}, Phys. Rev. B 62~(8)
  (2000) 4927--4944.

\bibitem{Prendergast2009}
D.~Prendergast, S.~G. Louie,
  \href{https://link.aps.org/doi/10.1103/PhysRevB.80.235126}{{Bloch-state-based
  interpolation: An efficient generalization of the Shirley approach to
  interpolating electronic structure}}, Phys. Rev. B 80 (2009) 235126.

\bibitem{RIM_method}
C.~A. Rozzi, D.~Varsano, A.~Marini, E.~K.~U. Gross, A.~Rubio,
  \href{https://link.aps.org/doi/10.1103/PhysRevB.73.205119}{Exact coulomb
  cutoff technique for supercell calculations}, Phys. Rev. B 73 (2006) 205119.

\bibitem{Bechstedt1992}
F.~Bechstedt, R.~Del~Sole, G.~Cappellini, L.~Reining,
  \href{https://www.sciencedirect.com/science/article/pii/003810989290476P?via%3Dihub}{An
  efficient method for calculating quasiparticle energies in semiconductors},
  Solid State Commun. 84~(7) (1992) 765--770.

\bibitem{Umari2010}
P.~Umari, G.~Stenuit, S.~Baroni,
  \href{https://journals.aps.org/prb/abstract/10.1103/PhysRevB.81.115104}{{GW
  quasiparticle spectra from occupied states only}}, Phys. Rev. B 81 (2010)
  115104.

\bibitem{Umari09}
P.~Umari, G.~Stenuit, S.~Baroni,
  \href{https://link.aps.org/doi/10.1103/PhysRevB.79.201104}{{Optimal
  representation of the polarization propagator for large-scale $GW$
  calculations}}, Phys. Rev. B 79 (2009) 201104.

\bibitem{Numerical_recipies}
W.~H. Press, S.~A. Teukolsky, W.~T. Vetterling, B.~P. Flannery,
  \href{http://www.cambridge.org/ch/academic/subjects/mathematics/numerical-recipes/numerical-recipes-art-scientific-computing-3rd-edition?format=HB&utm_source=shortlink&utm_medium=shortlink&utm_campaign=numericalrecipes}{Numerical
  Recipes: The Art of Scientific Computing}, 3rd Edition, Cambridge University
  Press, New York, NY, USA, 2007.

\bibitem{Benedict1999}
L.~Benedict, E.~Shirley,
  \href{https://journals.aps.org/prb/abstract/10.1103/PhysRevB.59.5441}{Ab
  initio calculation of ${\ensuremath{\epsilon}}_{2}(\ensuremath{\omega})$
  including the electron-hole interaction: Application to gan and
  ${\mathrm{caf}}_{2}$}, Phys. Rev. B 59~(8) (1999) 5441--5451.

\bibitem{Perdew1996}
J.~Perdew, K.~Burke, M.~Ernzerhof,
  \href{http://www.ncbi.nlm.nih.gov/pubmed/10062328}{{Generalized Gradient
  Approximation Made Simple.}}, Phys. Rev. Lett. 77~(18) (1996) 3865--3868.

\bibitem{Schlipf2015}
M.~Schlipf, F.~Gygi,
  \href{http://dx.doi.org/10.1016/j.cpc.2015.05.011}{{Optimization algorithm
  for the generation of ONCV pseudopotentials}}, Comput. Phys. Commun. 196
  (2015) 36--44.

\bibitem{MonkhorstPack}
H.~J. Monkhorst, J.~D. Pack,
  \href{https://link.aps.org/doi/10.1103/PhysRevB.13.5188}{{Special points for
  Brillouin-zone integrations}}, Phys. Rev. B 13 (1976) 5188--5192.

\bibitem{Pizzi2014}
G.~Pizzi, D.~Volja, B.~Kozinsky, M.~Fornari, N.~Marzari,
  \href{https://www.sciencedirect.com/science/article/pii/S0010465514001581?via%3Dihub}{{An
  updated version of BOLTZWANN: A code for the evaluation of thermoelectric and
  electronic transport properties with a maximally-localized Wannier functions
  basis}}, Comput. Phys. Commun. 185~(8) (2014) 2311--2312.

\bibitem{Mostofi2008}
A.~A. Mostofi, J.~R. Yates, Y.-S. Lee, I.~Souza, D.~Vanderbilt, N.~Marzari,
  \href{https://www.sciencedirect.com/science/article/pii/S0010465507004936?via%3Dihub}{{wannier90:
  A tool for obtaining maximally-localised Wannier functions}}, Comput. Phys.
  Commun. 178~(9) (2008) 685--699.

\bibitem{Marzari2012}
N.~Marzari, A.~A. Mostofi, J.~R. Yates, I.~Souza, D.~Vanderbilt,
  \href{https://link.aps.org/doi/10.1103/RevModPhys.84.1419}{Maximally
  localized wannier functions: Theory and applications}, Rev. Mod. Phys. 84
  (2012) 1419--1475.

\bibitem{Marzari1997}
N.~Marzari, D.~Vanderbilt,
  \href{https://link.aps.org/doi/10.1103/PhysRevB.56.12847}{Maximally localized
  generalized wannier functions for composite energy bands}, Phys. Rev. B 56
  (1997) 12847--12865.

\bibitem{Yates2007}
J.~R. Yates, X.~Wang, D.~Vanderbilt, I.~Souza,
  \href{https://link.aps.org/doi/10.1103/PhysRevB.75.195121}{Spectral and fermi
  surface properties from wannier interpolation}, Phys. Rev. B 75 (2007)
  195121.

\bibitem{Souza2001}
I.~Souza, N.~Marzari, D.~Vanderbilt,
  \href{https://link.aps.org/doi/10.1103/PhysRevB.65.035109}{Maximally
  localized wannier functions for entangled energy bands}, Phys. Rev. B 65
  (2001) 035109.

\bibitem{Damle2017}
A.~Damle, L.~Lin, L.~Ying,
  \href{http://www.sciencedirect.com/science/article/pii/S0021999116307215}{{SCDM-k:
  Localized orbitals for solids via selected columns of the density matrix}},
  J. Comput. Phys. 334 (2017) 1--15.

\bibitem{Damle2018}
A.~Damle, L.~Lin, \href{https://doi.org/10.1137/17M1129696}{{Disentanglement
  via Entanglement: A Unified Method for Wannier Localization}}, Multiscale
  Model. Simul. 16~(3) (2018) 1392--1410.

\bibitem{Albrecht98}
S.~Albrecht, L.~Reining, R.~Del~Sole, G.~Onida,
  \href{https://link.aps.org/doi/10.1103/PhysRevLett.80.4510}{Ab initio
  calculation of excitonic effects in the optical spectra of semiconductors},
  Phys. Rev. Lett. 80 (1998) 4510--4513.

\bibitem{Kammerlander12}
D.~Kammerlander, S.~Botti, M.~A.~L. Marques, A.~Marini, C.~Attaccalite,
  \href{https://link.aps.org/doi/10.1103/PhysRevB.86.125203}{{Speeding up the
  solution of the Bethe-Salpeter equation by a double-grid method and Wannier
  interpolation}}, Phys. Rev. B 86 (2012) 125203.

\bibitem{PZ81}
J.~P. Perdew, A.~Zunger,
  \href{https://link.aps.org/doi/10.1103/PhysRevB.23.5048}{Self-interaction
  correction to density-functional approximations for many-electron systems},
  Phys. Rev. B 23 (1981) 5048--5079.

\bibitem{albrecht_reply}
S.~Albrecht, L.~Reining, G.~Onida, V.~Olevano, R.~Del~Sole,
  \href{https://link.aps.org/doi/10.1103/PhysRevLett.83.3971}{Albrecht et al.
  reply:}, Phys. Rev. Lett. 83 (1999) 3971--3971.

\bibitem{Umari14}
P.~Umari, E.~Mosconi, F.~De~Angelis,
  \href{http://dx.doi.org/10.1038/srep04467}{{Relativistic GW calculations on
  CH3NH3PbI3 and CH3NH3SnI3 Perovskites for Solar Cell Applications}},
  Scientific Reports 4 (2014) 4467. 

\bibitem{Lautenschlager87}
P.~Lautenschlager, M.~Garriga, L.~Vina, M.~Cardona,
  \href{https://link.aps.org/doi/10.1103/PhysRevB.36.4821}{Temperature
  dependence of the dielectric function and interband critical points in
  silicon}, Phys. Rev. B 36 (1987) 4821--4830.

\end{thebibliography}

\newpage
\section*{Tables}

\begin{table}[H]
  \begin{center}
    \begin{tabular}{ll}
     \hline
     \textit{Input flag} & \textit{Description} \\
      \hline
         \texttt{prefix} & Same as in \texttt{pw.x} \\
          \texttt{outdir} & Same as in \texttt{pw.x} \\
\texttt{calc\_mode} & Calculation mode (0 for BSE, 1 for IP) \\
      \texttt{s\_bands}    & Threshold for the construction of the OB [bohr$^3$]\\
      \texttt{num\_nbndv}    & Occupied bands\\
      \texttt{num\_val}    & Valence bands (starting from HOMO)\\
      \texttt{num\_cond}    & Conduction bands (starting from LUMO)\\
      \texttt{nkpoints(1)} & k-grid ($n_1$) \\
      \texttt{nkpoints(2)} & k-grid ($n_2$) \\
      \texttt{nkpoints(3)} & k-grid ($n_3$) \\
      \texttt{s\_product} & Threshold for the construction of the OPB [bohr$^3$] (BSE)\\
      \texttt{w\_type} & Screened Coulomb interaction from \texttt{GWL} calculation (0) or \\
      & from model microscopic dielectric function (1) (BSE)\\       
      \texttt{nonlocal\_commutator} & Flag to calculate the non-local commutator (IP)\\
      \texttt{epsm} & Macroscopic dielectric constant $\epsilon_m$ from model microscopic \\
      & dielectric function (if \texttt{w\_type}=1) (BSE)\\
      \texttt{lambdam} & Screening length $\overline{\lambda}$ from model microscopic \\
      & dielectric function (if \texttt{w\_type}=1) [bohr$^{-1}$] (BSE)\\
      \texttt{numpw} & Dimension of polarizability basis (if \texttt{w\_type}=0) (BSE)\\
      \texttt{l\_truncated\_coulomb} & Flag to activate truncation of Coulomb \\
      & interaction (BSE)\\
      \texttt{truncation\_radius} & Radius at which Coloumb interaction \\
      & is truncated  [bohr] (BSE)\\
      \hline
    \end{tabular}
    \caption{Input keywords for \texttt{simple.x}. The input file must contain the string \texttt{\&inputsimple} before inserting any input flag.} \label{tab: input_simple}
  \end{center}
\end{table}

\begin{table}[H]
  \begin{center}
    \begin{tabular}{ll}
     \hline
     \textit{Input flag} & \textit{Description} \\
      \hline
         \texttt{simpleip\_in\%}\texttt{prefix} & Same as in \texttt{pw.x} \\
            \texttt{simpleip\_in\%}\texttt{outdir} & Same as in \texttt{pw.x} \\      
      \texttt{simpleip\_in\%}\texttt{interp\_grid(1)} & Interpolation k-grid ($m_1$) \\
      \texttt{simpleip\_in\%}\texttt{interp\_grid(2)} & Interpolation k-grid ($m_2$) \\
      \texttt{simpleip\_in\%}\texttt{interp\_grid(3)} & Interpolation k-grid ($m_3$) \\
      \texttt{simpleip\_in\%}\texttt{fermi\_ngauss} & Broadening type for Fermi energy (integer $n$) \\
      & \begin{footnotesize}  $n=-99$ Fermi-Dirac,   $n=-1$ Marzari-Vanderbilt, \end{footnotesize} \\
      & \begin{footnotesize} $n \geq 0$ Methfessel-Paxton \end{footnotesize}\\
      \texttt{simpleip\_in\%}\texttt{fermi\_degauss} & Broadening for Fermi energy [Ry]\\
      \texttt{simpleip\_in\%}\texttt{drude\_degauss} & Broadening for Drude plasma frequency [Ry]\\
      \texttt{simpleip\_in\%}\texttt{inter\_broadening} & Empirical broadening for $\varepsilon_{\text{IP}}^{\text{inter}}(\hat{\mb{q}},\omega)$ [Ry]\\
      \texttt{simpleip\_in\%}\texttt{intra\_broadening} & Empirical broadening for $\varepsilon_{\text{IP}}^{\text{intra}}(\hat{\mb{q}},\omega)$ [Ry]\\
      \texttt{simpleip\_in\%}\texttt{wmin} & Minimum frequency for $\varepsilon_{\text{IP}}(\hat{\mb{q}},\omega)$ [Ry]\\
      \texttt{simpleip\_in\%}\texttt{wmax} & Maximum frequency for $\varepsilon_{\text{IP}}(\hat{\mb{q}},\omega)$ [Ry]\\
      \texttt{simpleip\_in\%}\texttt{nw} & Number of frequencies in the range [\texttt{wmin}, \texttt{wmax}]\\
      \texttt{simpleip\_in\%}\texttt{nonlocal\_commutator} & Flag to include the non-local commutator\\
      \texttt{simpleip\_in\%}\texttt{nonlocal\_interpolation} & Flag to use linear interpolation of non-local \\ & contributions (experimental feature)\\
      \hline
    \end{tabular}
    \caption{Input keywords for \texttt{simple\_ip.x}. The input file must contain the string \texttt{\&inputsimpleip} before inserting any input flag. The prefix of each input flag is \texttt{simpleip\_in\%} as reported in the table.}\label{tab: input_simpleip}
  \end{center}
\end{table}

\begin{table}[H]
  \begin{center}
    \begin{tabular}{ll}
     \hline
     \textit{Input flag} & \textit{Description} \\
      \hline
         \texttt{simple\_in\%}\texttt{prefix} & Same as in \texttt{pw.x} \\
            \texttt{simple\_in\%}\texttt{outdir} & Same as in \texttt{pw.x} \\
              \texttt{simple\_in\%}\texttt{scissor} & Rigid scissor energy for opening DFT gap [eV] \\
                \texttt{simple\_in\%}\texttt{spin\_state} & In case of collinear spin calculation selects triplet (0) or \\
                   &or singlet (1) excitons , (2) for the case of non-collinear spin \\
              \texttt{simple\_in\%}\texttt{h\_level} & Terms in  excitonic Hamiltonian: (0) only diagonal,  (1) as (0)\\
              & plus exchange term, (2) as (1) plus direct term of bare \\
              & Coulomb  part, (3) as (2) plus direct term of correlation part  \\
             \texttt{simple\_in\%}\texttt{task} &If set to 0  find the lowest excitons, if  set to 1 find the\\
             & $\alpha_i$ and $\beta_i$ coefficients of the Haydock recursive method \\
              \texttt{simple\_in\%}\texttt{nvec} & Number of excitons to be found\\
               \texttt{simple\_in\%}\texttt{thr\_evc} & Threshold for defining converged excitonic eigenenergies [eV]\\
                 \texttt{simple\_in\%}\texttt{diago} & Algorithm for finding excitonic eigenvectors:\\
                 &(0) steepest descent (1) conjugate gradient  \\
                  \texttt{simple\_in\%}\texttt{max\_nstep} & Maximum number of steps during minimizations \\
                   \texttt{simple\_in\%}\texttt{lanczos\_step} & Number of steps for the Haydock recursive method\\
       \hline
    \end{tabular}
    \caption{Input keywords for \texttt{simple\_bse.x}. The input file must contain the string \texttt{\&inputsimple} before inserting any input flag. The prefix of each input flag is \texttt{simple\_in\%} as reported in the table.}\label{tab: input_simplebse}
  \end{center}
\end{table}

\section*{Figures}

\begin{figure}[hbtp]
\centering
\includegraphics[width=\textwidth]{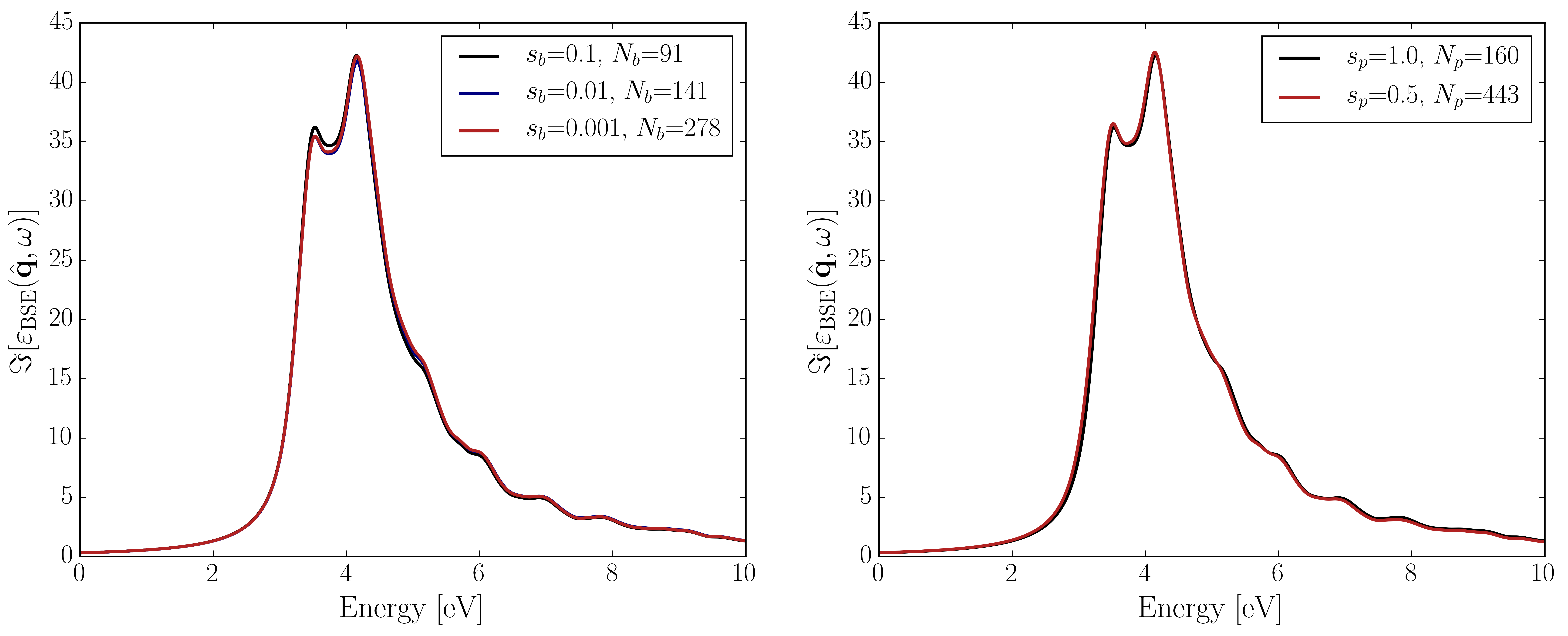}
\caption{Convergence of the imaginary part of the BSE complex dielectric function $\Im{\left[\varepsilon_{\text{BSE}}(\hat{\mb{q}},\omega)\right]}$ for bulk Si with respect to the $s_b$ threshold (in bohr$^3$) while the $s_p$ threshold is kept fixed to $1.0$ bohr$^3$ (left panel) and with respect to the $s_p$ threshold (in bohr$^3$) while the $s_b$ threshold is kept fixed to $0.1$ bohr$^3$ (right panel). The corresponding dimensions $N_b$ and $N_p$ of the optimal basis and of the optimal product basis, respectively, are also reported.  \label{fig: BSE_sb-sp_convergence}}
\end{figure}

\end{document}